\documentclass[12pt]{article}

\usepackage{amsmath,amssymb,amsfonts}
\usepackage{psfrag}
\usepackage{color}
\definecolor{darkblue}{rgb}{0.1,0.1,.7}
\usepackage[colorlinks, linkcolor=darkblue, citecolor=darkblue, urlcolor=darkblue, linktocpage]{hyperref} 
\usepackage[square, comma, compress,numbers]{natbib}
\usepackage[]{amsmath}
\usepackage[]{graphicx}
\usepackage[]{latexsym}
\usepackage{geometry}
\usepackage{amscd}
\usepackage[all,cmtip]{xy}
\usepackage{mathrsfs}
\usepackage{bbold}
\usepackage{subfigure}
\usepackage[margin=10pt,font=small,labelfont=bf]{caption}
\usepackage{dsdshorthand}
\usepackage{simplewick}
\usepackage{changepage}
\numberwithin{equation}{section}

\renewcommand{\be}{\begin{eqnarray}}
\renewcommand{\ee}{\end{eqnarray}}
\newcommand{\bea}{\begin{eqnarray}}
\newcommand{\eea}{\end{eqvnarray}}

\def\beq{\begin{equation}} 
\def\eeq{\end{equation}} 
\def\<{\langle}
\def\>{\rangle}

\def\nn{\nonumber} 
 
\def\cO {{\cal O}}

\def\cN {{\cal N}}

\begin{document}

\vspace*{-.6in} \thispagestyle{empty}
\begin{flushright}
CERN-PH-TH-2015-097
\end{flushright}
\vspace{.2in} {\Large
\begin{center}
{\bf Bootstrapping the $O(N)$ Archipelago}
\end{center}
}
\vspace{.2in}
\begin{center}
{\bf 
Filip Kos$^{a}$, 
David Poland$^{a}$,
David Simmons-Duffin$^{b}$,
Alessandro Vichi$^{c}$} 
\\
\vspace{.2in} 
$^a$ {\it  Department of Physics, Yale University, New Haven, CT 06520}\\
$^b$ {\it School of Natural Sciences, Institute for Advanced Study, Princeton, New Jersey 08540}\\
$^c$ {\it Theory Division, CERN, Geneva, Switzerland}
\end{center}

\vspace{.2in}

\begin{abstract}
We study 3d CFTs with an $O(N)$ global symmetry using the conformal bootstrap for a system of mixed correlators. Specifically, we consider all nonvanishing scalar four-point functions containing the lowest dimension $O(N)$ vector $\phi_i$ and the lowest dimension $O(N)$ singlet $s$, assumed to be the only relevant operators in their symmetry representations. The constraints of crossing symmetry and unitarity for these four-point functions force the scaling dimensions $(\Delta_\phi, \Delta_s)$ to lie inside small islands. We also make rigorous determinations of current two-point functions in the $O(2)$ and $O(3)$ models, with applications to transport in condensed matter systems.
\end{abstract}

\newpage

\tableofcontents

\newpage


\section{Introduction}

Conformal field theories (CFTs) lie at the heart of theoretical physics, describing critical phenomena in statistical and condensed matter systems, quantum gravity via the AdS/CFT correspondence, and possible solutions to the hierarchy problem (and other puzzles) in physics beyond the standard model.  Quite generally, they serve as the endpoints of renormalization group flows in quantum field theory. The conformal bootstrap~\cite{Ferrara:1973yt,Polyakov:1974gs} aims to use general consistency conditions to map out and solve CFTs, even when they are strongly-coupled and do not have a useful Lagrangian description. 

In recent years great progress has been made in the conformal bootstrap in $d>2$, including rigorous bounds on operator dimensions and operator product expansion (OPE) coefficients~\cite{Rattazzi:2008pe,Rychkov:2009ij,Caracciolo:2009bx,Poland:2010wg,Rattazzi:2010gj,Rattazzi:2010yc,Vichi:2011ux,Poland:2011ey,Rychkov:2011et,ElShowk:2012ht,Liendo:2012hy,Beem:2013qxa,Kos:2013tga,El-Showk:2013nia,Alday:2013opa,Gaiotto:2013nva,Bashkirov:2013vya,Berkooz:2014yda,Nakayama:2014lva,Nakayama:2014yia,Alday:2014qfa,Chester:2014fya,Caracciolo:2014cxa,Nakayama:2014sba,Paulos:2014vya,Bae:2014hia,Beem:2014zpa,Chester:2014gqa,Bobev:2015vsa,Bobev:2015jxa}, analytical constraints~\cite{Heemskerk:2009pn,Heemskerk:2010ty,Fitzpatrick:2012yx,Komargodski:2012ek,Beem:2013sza,Fitzpatrick:2014vua,Beem:2014kka,Alday:2014tsa,Chester:2014mea,Kaviraj:2015cxa,Alday:2015eya,Kaviraj:2015xsa,Fitzpatrick:2015qma}, and methods for approximate direct solutions to the bootstrap~\cite{ElShowk:2012hu,Gliozzi:2013ysa,Gliozzi:2014jsa,Gliozzi:2015qsa}, including a precise determination of the low-lying spectrum in the 3d Ising model under the conjecture that the conformal central charge is minimized~\cite{El-Showk:2014dwa}. These results have come almost exclusively from analyzing 4-point correlation functions of identical operators. It is tantalizing that even more powerful constraints may come from mixed correlators.

In~\cite{Kos:2014bka} some of the present authors demonstrated that semidefinite programming techniques can very generally be applied to systems of mixed correlators. In 3d CFTs with a $\mathbb{Z}_2$ symmetry, one relevant $\mathbb{Z}_2$-odd operator $\sigma$, and one relevant $\mathbb{Z}_2$-even operator $\epsilon$, the mixed correlator bootstrap leads to a small and {\it isolated} allowed region in operator dimension space consistent with the known dimensions in the 3d Ising CFT. With the assistance of improved algorithms for high-precision semidefinite programming~\cite{Simmons-Duffin:2015qma}, this approach has culminated in the world's most precise determinations of the leading operator dimensions $(\Delta_{\sigma}, \Delta_{\epsilon}) = (0.518151(6),1.41264(6))$ in the 3d Ising CFT.

The immediate question is whether the same approach can be used to rigorously isolate and precisely determine spectra in the zoo of other known (and perhaps unknown) CFTs, particularly those with physical importance. In this work we focus on 3d CFTs with $O(N)$ global symmetry, previously studied using numerical bootstrap techniques in~\cite{Kos:2013tga,Nakayama:2014yia}. We will show that the CFTs known as the $O(N)$ vector models can be similarly isolated using a system of mixed correlators containing the leading $O(N)$ vector $\phi_i$ and singlet $s$, assumed to be the only relevant operators in their symmetry representations. 

\begin{figure}[htbp]
\begin{center}
\includegraphics[scale=1.3]{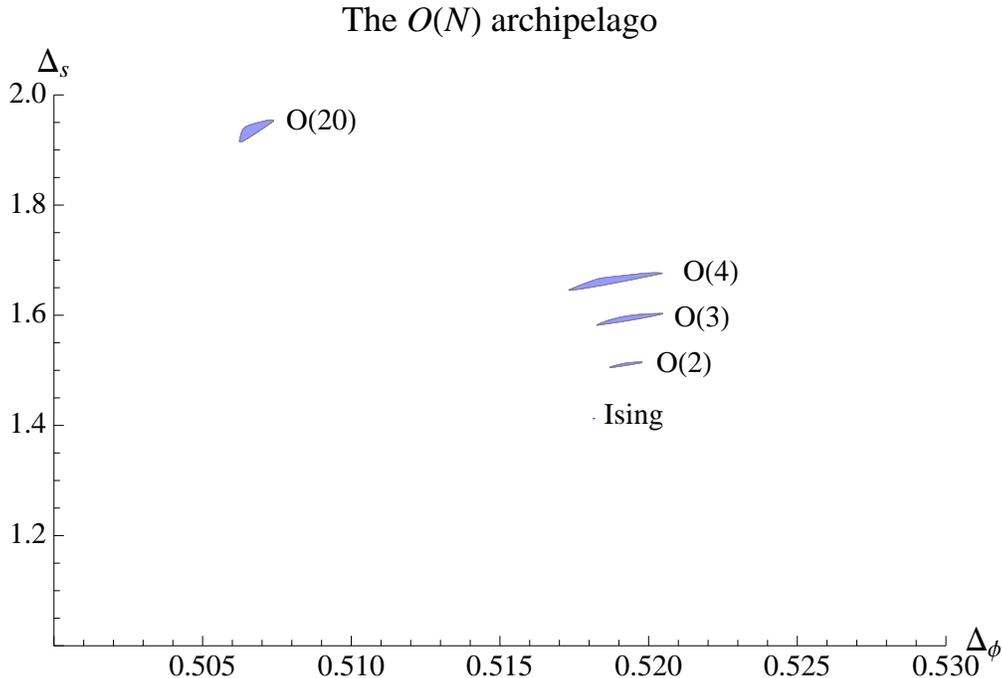}
\caption{Allowed regions for operator dimensions in 3d CFTs with an $O(N)$ global symmetry and exactly one relevant scalar $\phi_i$ in the vector representation and one relevant scalar $s$ in the singlet representation of $O(N)$, for $N=1,2,3,4,20$.  The case $N=1$, corresponding to the 3d Ising model, is from~\cite{Kos:2014bka}. The allowed regions for $N=2,3,4,20$ were computed with $\Lambda=35$, where $\Lambda$ (defined in appendix~\ref{sec:appA}) is related to the number of derivatives of the crossing equation used. Each region is roughly triangular, with an upper-left vertex that corresponds to the kinks in previous bounds \cite{Kos:2013tga}. Further allowed regions may exist outside the range of this plot; we leave their exploration to future work. }
\label{fig:Archipelago}
\end{center}
\end{figure}

We focus on the physically most interesting cases $N=2,3,4$ (e.g., see~\cite{Pelissetto:2000ek}) where the large-$N$ expansion fails. We do additional checks at $N=20$. A summary of the constraints on the leading scaling dimensions found in this work are shown in figure~\ref{fig:Archipelago}. We also make precise determinations of the current central charge $\<J J\> \propto C_J$ for $N=2,3$. This coefficient is particularly interesting because it describes conductivity properties of materials in the vicinity of their critical point~\cite{Katz:2014rla}.
 
The 3d $O(2)$ model (or $XY$ model) has a beautiful experimental realization in superfluid ${}^4$He~\cite{Lipa:2003zz} which has yielded results for $\Delta_{s}$ that are in $\sim8\sigma$ tension with the leading Monte Carlo and high temperature expansion computations~\cite{Campostrini:2006ms}. Our results are not yet precise enough to resolve this discrepancy, but we are optimistic that the approach we outline in this work will be able to do so in the near future. More generally, the results of this work give us hope that the same techniques can be used to to solve other interesting strongly-coupled CFTs, such as the 3d Gross-Neveu models, 3d Chern-Simons and gauge theories coupled to matter, 4d QCD in the conformal window, $\cN=4$ supersymmetric Yang-Mills theory, and more.

The structure of this paper is as follows. In section~\ref{sec:crossing}, we summarize the crossing symmetry conditions arising from systems of correlators in 3d CFTs with $O(N)$ symmetry, and discuss how to study them with semidefinite programming. In section~\ref{sec:results}, we describe our results and in section~\ref{sec:conclusions} we discuss several directions for future work. Details of our implementation are given in appendix~\ref{sec:appA}. An exploration of the role of the leading symmetric tensor is given in appendix~\ref{sec:appB}.

\section{Crossing Symmetry with Multiple Correlators}
\label{sec:crossing}
Let us begin by summarizing the general form of the crossing relation for a collection of scalar fields $\phi_i = (\phi_1,\phi_2,\phi_3,\ldots)$. We take the $\phi_i$ to have dimensions $\Delta_i$ and for the moment we do not assume any symmetry relating them. Taking the OPE of the first two and last two operators, the 4-point function looks like:
\be
\<\phi_i(x_1) \phi_j(x_2) \phi_k(x_3) \phi_l(x_4) \> &=& \frac{1}{x_{12}^{\De_i + \De_j} x_{34}^{\De_k+\De_l}} \left(\frac{x_{24}}{x_{14}}\right)^{\De_{ij}} \left(\frac{x_{14}}{x_{13}}\right)^{\De_{kl}} \sum_{\cO} \l_{ij\cO} \l_{kl\cO} g_{\De,\ell}^{\De_{ij},\De_{kl}}(u,v), \nonumber\\
u &=& \frac{x_{12}^2x_{34}^2}{x_{13}^2x_{24}^2},\quad
v = \frac{x_{14}^2x_{23}^2}{x_{13}^2x_{24}^2},
\ee
where $x_{ij} \equiv |x_i - x_j|$, $\De_{ij} \equiv \De_i - \De_j$, and $u$, $v$ are the standard conformal invariants. The subscripts $\Delta,\ell$ refer to the dimension and spin of the operator $\cO$.  We refer to~\cite{Kos:2014bka} for details about how to compute the conformal blocks $g_{\De,\ell}^{\De_{ij},\De_{kl}}(u,v)$ in any dimension and for arbitrary values of $\De_{ij} $.   
We also have the symmetry property $\l_{ij\cO} = (-1)^{\ell} \l_{ji\cO}$.

Crossing symmetry of the correlation function requires that OPEs taken in different orders must produce the same result. As an example, exchanging $(1,i) \leftrightarrow (3,k)$ gives the conditions:
\be\label{eq:crossing}
v^{\frac{\De_k+\De_j}{2}} \sum_{\cO} \l_{ij\cO} \l_{kl\cO} g_{\De,\ell}^{\De_{ij},\De_{kl}}(u,v) = u^{\frac{\De_i+\De_j}{2}} \sum_{\cO} \l_{kj\cO} \l_{il\cO} g_{\De,\ell}^{\De_{kj},\De_{il}}(v,u).
\ee
It is convenient to symmetrize/anti-symmetrize in $u,v$, which leads to the two equations:
\be\label{eq:crossingsym}
0 &=& \sum_{\cO} \left[ \l_{ij\cO} \l_{kl\cO} F^{ij,kl}_{\mp,\De,\ell}(u,v) \pm \l_{kj\cO} \l_{il\cO} F^{kj,il}_{\mp,\De,\ell}(u,v) \right],
\ee
where
\be
F_{\mp,\De,\ell}^{ij,kl}(u,v)  \equiv v^{\frac{\De_k+\De_j}{2}} g_{\De,\ell}^{\De_{ij},\De_{kl}}(u,v) \mp  u^{\frac{\De_k+\De_j}{2}} g_{\De,\ell}^{\De_{ij},\De_{kl}}(v,u).
\ee
The functions $F_{\mp,\De,\ell}^{ij,kl}$ are symmetric under exchanging $i \leftrightarrow k$ and $j \leftrightarrow l$.

\subsection{$O(N)$ Models}
\label{sec:oncrossingeqs}

We now restrict our discussion to the case where $\phi_i$ transforms in the vector representation of a global $O(N)$ symmetry.
When the fields entering the four-point function are charged under global symmetries, the conformal block expansion can be organized in symmetry structures corresponding to irreducible representations appearing in the OPE $\phi_i\times\phi_j$. This gives the equations\footnote{Note that we are following the conformal block conventions of~\cite{Kos:2014bka}, which contain a factor of $(-1)^{\ell}$ relative to the conventions used in the previous global symmetry studies~\cite{Poland:2011ey,Kos:2013tga}. This leads to a different sign in front of the contributions of the $\cO_A$ operators.} 
\be\label{eq:ON}
0 &=& (\de_{ij}\de_{kl} \pm \de_{jk}\de_{il})\sum_{\cO_S,\ell^+} \l_{\phi\phi \cO_S}^2F_{\mp,\De,\ell}^{\f\f,\f\f}\nn\\
&&+\p{\p{\delta_{ik}\delta_{jl}+\delta_{il}\delta_{jk}-\frac2N\delta_{ij}\delta_{kl}} \pm \p{\delta_{ik}\delta_{jl}+\delta_{ij}\delta_{kl}-\frac2N\delta_{jk}\delta_{il}}}\sum_{\cO_T,\ell^+} \l_{\f\f \cO_T}^2 F_{\mp,\De,\ell}^{\f\f,\f\f}\nn\\
&&+\p{(\delta_{ik}\delta_{jl} - \delta_{il}\delta_{jk})\pm (\delta_{ik}\delta_{jl}  - \delta_{ij}\delta_{kl})}
\sum_{\cO_A,\ell^-}\l_{\f\f \cO_A}^2 F_{\mp,\De,\ell}^{\f\f,\f\f},
\ee
which lead to three independent sum rules after reading off the coefficients of each index structure. Here, $\cO_S,\cO_T,\cO_A$ denote operators in the singlet, traceless symmetric tensor, and antisymmetric tensor representations of $O(N)$, $\ell^+$ refers to operators with even spin, and $\ell^{-}$ refers to odd spin. The sum over spins is determined by the symmetry properties of the representations under exchange of two indices.

In what follows, we will use $s,s',s'',\dots$ to refer to the singlet scalars in increasing order of dimension. For example, $s$ is the lowest-dimension singlet scalar in the theory. Similarly, $t,t',t'',\dots$ and $\f,\f',\f'',\dots$ refer to scalars in the traceless symmetric tensor and vector representations, in increasing order of dimension.

We would like to supplement the above equations with crossing symmetry constraints from other four-point functions. The simplest choice is to consider all nonvanishing four-point functions of $\f_i$ with the lowest dimension singlet scalar operator $s$.  Another interesting choice would be the lowest dimension scalar in the traceless symmetric tensor representation $t_{ij}$. However the OPEs $t_{ij}\x t_{kl}$ and $t_{ij}\x \f_k$ contain many additional $O(N)$ representations, increasing the complexity of the crossing equations.  We leave the analysis of external $t_{ij}$ operators to the future.

Thus we consider the four-point functions $\<\phi_i\phi_j s s\>$ and $\<ssss\>$, which give rise to four additional sum rules after grouping the terms with the same index structure. In total this leads to a system of seven equations:
\be
0&=&\sum_{\cO_T,\ell^+} \l_{\f\f \cO_T}^2 F^{\phi\phi,\phi\phi}_{-,\De,\ell} + \sum_{\cO_A,\ell^-} \l_{\f\f \cO_A}^2 F^{\phi\phi,\phi\phi}_{-,\De,\ell},\nn\\
0&=&\sum_{\cO_S,\ell^+} \l_{\f\f \cO_S}^2 F^{\phi\phi,\phi\phi}_{-,\De,\ell} + (1-\frac2N)\sum_{\cO_T,\ell^+} \l_{\f\f \cO_T}^2 F^{\phi\phi,\phi\phi}_{-,\De,\ell} - \sum_{\cO_A,\ell^-} \l_{\f\f \cO_A}^2 F^{\phi\phi,\phi\phi}_{-,\De,\ell},\nn\\
0&=&\sum_{\cO_S,\ell^+} \l_{\f\f \cO_S}^2 F^{\phi\phi,\phi\phi}_{+,\De,\ell} - (1+\frac2N)\sum_{\cO_T,\ell^+} \l_{\f\f \cO_T}^2 F^{\phi\phi,\phi\phi}_{+,\De,\ell} + \sum_{\cO_A,\ell^-} \l_{\f\f \cO_A}^2 F^{\phi\phi,\phi\phi}_{+,\De,\ell},\nn\\
0&=&\sum_{\cO_S,\ell^+} \l_{ss \cO_S}^2 F^{ss,ss}_{-,\De,\ell},\nn\\
0&=&\sum_{\cO_V,\ell^\pm} \l_{\f s \cO_V}^2 F^{\phi s,\phi s}_{-,\De,\ell},\nn\\
0&=& \sum_{\cO_S,\ell^+} \l_{\f\f \cO_S} \l_{ss \cO_S} F^{\phi\phi,ss}_{\mp,\De,\ell} \pm \sum_{\cO_V,\ell^\pm} (-1)^\ell \l_{\f s \cO_V}^2 F^{s\phi,\phi s}_{\mp,\De,\ell}.
\ee
Note that the final line represents two equations, corresponding to the choice of $\pm$. We can rewrite these equations in vector notation as
\be\label{eq:vectoreq}
0 &=& \sum_{\cO_S,\ell^+} \left(\begin{array}{ccc} \l_{\f\f \cO_S} & \l_{ss \cO_S} \end{array} \right) \vec{V}_{S,\De,\ell} \left( \begin{array}{c} \l_{\f\f \cO_S} \\ \l_{ss \cO_S} \end{array} \right) +\sum_{\cO_T,\ell^+} \l_{\f\f \cO_T}^2 \vec{V}_{T,\De,\ell}\nn\\
&&+ \sum_{\cO_A,\ell^-} \l_{\f\f \cO_A}^2 \vec{V}_{A,\De,\ell}+ \sum_{\cO_V,\ell^\pm} \l_{\f s \cO_V}^2 \vec{V}_{V,\De,\ell},
\ee
where $\vec{V}_{T}, \vec{V}_A, \vec{V}_V$ are a 7-dimensional vectors and $\vec{V}_{S}$ is a 7-vector of $2 \times 2$ matrices:

\be
\vec{V}_{T,\De,\ell} = \left( \begin{array}{c} F_{-,\De,\ell}^{\phi\phi,\phi\phi}  \\ \left(1-\frac2N\right)F_{-,\De,\ell}^{\phi\phi,\phi\phi} \\ -\left(1+\frac2N\right)F_{+,\De,\ell}^{\phi\phi,\phi\phi}\\0 \\ 0 \\ 0\\0 \end{array} \right),
\ 
\vec{V}_{A,\De,\ell} = \left( \begin{array}{c} F_{-,\De,\ell}^{\phi\phi,\phi\phi}  \\ - F_{-,\De,\ell}^{\phi\phi,\phi\phi} \\ F_{+,\De,\ell}^{\phi\phi,\phi\phi}\\0 \\ 0 \\ 0\\0 \end{array} \right),
\ 
\vec{V}_{V,\De,\ell} = \left( \begin{array}{c} 0  \\ 0 \\ 0 \\0 \\ F_{-,\De,\ell}^{\phi s,\phi s}\\ (-1)^\ell F_{-,\De,\ell}^{s\phi,\phi s} \\-(-1)^\ell F_{+,\De,\ell}^{s\phi,\phi s}  \end{array} \right),\nn\\
\ee

\be
\vec{V}_{S,\De,\ell} = \left( \begin{array}{c} 
 \left( \begin{array}{cc} 0 & 0 \\ 0 & 0 \end{array} \right) \\
\left( \begin{array}{cc} F^{\phi\phi,\phi\phi}_{-,\De,\ell}(u,v) & 0 \\ 0 & 0 \end{array} \right) \\
\left( \begin{array}{cc} F^{\phi\phi,\phi\phi}_{+,\De,\ell}(u,v) & 0 \\ 0 & 0 \end{array} \right)\\
 \left( \begin{array}{cc} 0 & 0 \\ 0 & F^{ss,ss}_{-,\De,\ell}(u,v) \end{array} \right) \\
\left( \begin{array}{cc} 0 & 0 \\ 0 & 0 \end{array} \right) \\
 \left( \begin{array}{cc} 0 & \frac12 F^{\phi\phi,ss}_{-,\De,\ell}(u,v) \\ \frac12 F^{\phi\phi,ss}_{-,\De,\ell}(u,v) & 0 \end{array} \right) \\
 \left( \begin{array}{cc} 0 & \frac12 F^{\phi\phi,ss}_{+,\De,\ell}(u,v) \\ \frac12 F^{\phi\phi,ss}_{+,\De,\ell}(u,v) & 0 \end{array} \right) \end{array} \right).
\ee

\subsubsection{A Note on Symmetries}

We are primarily interested in theories with $O(N)$ symmetry. However, our bounds will also apply to theories with the weaker condition of $SO(N)$ symmetry. This point deserves discussion.

The group $O(N)$ includes reflections, so its representation theory is slightly different from that of $SO(N)$. In particular  $\epsilon_{i_1\dots i_N}$ is not an invariant tensor of $O(N)$ because it changes sign under reflections.  For odd $N=2k+1$,  $O(2k+1)$ symmetry is equivalent to $SO(2k+1)$ symmetry plus an additional $\Z_2$ symmetry.  For even $N=2k$, the orthogonal group is a semidirect product $O(2k)\cong \Z_2 \ltimes \SO(2k)$, so it is not equivalent to an extra $\Z_2$.

Let us consider whether the crossing equations must be modified in the case of only $SO(N)$ symmetry.  In theories with $SO(2)$ symmetry, the antisymmetric tensor representation is isomorphic to the singlet representation. (This is not true for $O(2)$ because the isomorphism involves $\epsilon_{ij}$.)  However in the crossing equation (\ref{eq:vectoreq}), antisymmetric tensors appear with odd spin, while singlets appear with even spin.  Thus, the coincidence between $A$ and $S$ does not lead to additional relations in (\ref{eq:vectoreq}).

For theories with $SO(3)$ symmetry, the antisymmetric tensor representation is equivalent to the vector representation.  Thus, antisymmetric odd spin operators appearing in $\f\x\f$ may also appear in $\f \x s$.  This does not affect (\ref{eq:vectoreq}) because there is no a priori relationship between $\l_{\f\f\cO}$ and $\l_{\f s\cO}$.  However, it is now possible to have a nonvanishing four-point function $\<\f_i\f_j\f_k s\>$ proportional to $\e_{ijk}$.  Including crossing symmetry of this four-point function cannot change the resulting dimension bounds without additional assumptions.  The reason is as follows.  Any bound computed from (\ref{eq:vectoreq}) without using crossing of $\<\f\f\f s\>$ is still valid.  Hence, the bounds cannot weaken.  However, because any $O(3)$-invariant theory is also $SO(3)$-invariant, any bound computed while demanding crossing of $\<\f\f\f s\>$ must also apply to $O(3)$-invariant theories.  So the bounds cannot strengthen.  Crossing for $\<\f\f\f s\>$ only becomes important if we input that $\l_{\f\f\cO}\l_{\f s\cO}$ is nonzero for a particular operator.\footnote{In practice, this means we would group this operator with the unit operator and other operators whose OPE coefficients are known in the semidefinite program.} This would guarantee our theory does not have $O(3)$ symmetry.

For $SO(4)$, the new ingredient is that the antisymmetric tensor representation can be decomposed into self-dual and anti-self-dual two-forms. As explained in \cite{Poland:2011ey}, this leads to an additional independent sum rule
\be
\sum_{A_+,\ell^-} \lambda_{\f\f \cO_{A_+}}^2 F^{\f\f;\f\f}_{\De,\ell} - \sum_{A_-,\ell^-}\l_{\f\f \cO_{A_-}}^2 F^{\f\f;\f\f}_{\De,\ell} &=& 0,
\label{eq:selfantiselfdual}
\ee
where $A_\pm$ represent self-dual and anti-self-dual operators.  By the same reasoning as in the case of $SO(3)$, this crossing equation cannot affect the bounds from (\ref{eq:vectoreq}) without additional assumptions.  We can also see this directly from (\ref{eq:selfantiselfdual}) together with (\ref{eq:vectoreq}): in the semidefinite program used to derive operator dimension bounds, we may always take the functional acting on (\ref{eq:selfantiselfdual}) to be zero.  An exception occurs if we know an operator is present with $\l_{\f\f \cO_{A_+}}\neq 0$ but $\l_{\f\f \cO_{A_-}}=0$ (or vice versa).
Then we can include that operator with other operators whose OPE coefficients are known (usually just the unit operator) and the resulting semidefinite program will be different.

For $SO(N)$ with $N\geq 5$, no coincidences occur in the representation ring that would be relevant for the system of correlators considered here. In conclusion, (\ref{eq:vectoreq}) and the semidefinite program discussed below remain valid in the case of $SO(N)$ symmetry. Bounds on theories with $SO(N)$ symmetry can differ only if we input additional information into the crossing equations that distinguishes them from $O(N)$-invariant theories (for example, known nonzero OPE coefficients).

\subsection{Bounds from Semidefinite Programming}

As explained in \cite{Kos:2014bka}, solutions to vector equations of the form (\ref{eq:vectoreq}) can be constrained using semidefinite programming (SDP).  We refer to \cite{Kos:2014bka} for details. Here we simply present the problem we must solve.  To rule out a hypothetical CFT spectrum, we must find a vector of linear functionals $\vec\alpha = (\alpha_1,\alpha_2,...,\alpha_7)$ such that
\be\label{eq:functionalconditions}
& \left(\begin{array}{ccc} 1 & 1 \end{array}\right) \vec\alpha\cdot \vec{V}_{S,0,0} \left( \begin{array}{c} 1 \\ 1 \end{array} \right) \geq 0,&\text{for the identity operator},\\
&\vec\alpha\cdot \vec{V}_{T,\De,\ell} \geq 0,&\text{for all traceless symetric tensors with $\ell$ even},\label{eq:functional-symtensor} \\
&\vec\alpha\cdot \vec{V}_{A,\De,\ell}  \geq 0,&\text{for all antisymmetric tensors with $\ell$ odd},\\
&\vec\alpha\cdot \vec{V}_{V,\De,\ell} \geq 0,&\text{for all $O(N)$ vectors with any $\ell$}\label{eq:functional-fundamental},\\
&\vec\alpha\cdot \vec{V}_{S,\De,\ell} \succeq 0,&\text{for all singlets with $\ell$ even}\label{eq:functional-singlet}.
\ee
Here, the notation ``$\succeq 0$" means ``is positive semidefinite."
If such a functional exists for a hypothetical CFT spectrum, then that spectrum is inconsistent with crossing symmetry. In addition to any explicit assumptions placed on the allowed values of $\Delta$, we impose that all operators must satisfy the unitarity bound
\be
\De &\geq& \left\{\begin{array}{ll}
\ell + D - 2 & \ell > 0\\
\frac{D-2}{2} & \ell = 0
\end{array}
\right.,
\ee
where $D=3$ is the spacetime dimension.

Additional information about the spectrum can weaken the above constraints, making the search for the functional $\vec\alpha$ easier, and further restricting the allowed theories. A few specific assumptions will be important in what follows:
\begin{itemize}
\item The 3d $O(N)$ vector models, which are our main focus, are believed to have exactly one relevant singlet scalar $s$, $O(N)$ vector scalar $\f_i$, and traceless symmetric scalar $t_{ij}$.\footnote{Additional relevant scalars could be present in other representations.}  We will often assume gaps to the second-lowest dimension operators $s', \f_i', t_{ij}'$ in each of these sectors. These assumptions affect (\ref{eq:functional-symtensor}), (\ref{eq:functional-fundamental}), and (\ref{eq:functional-singlet}).

\item Another important input is the equality of the OPE coefficients $\l_{\phi\phi s}=\l_{\phi s\phi}$. This is a trivial consequence of conformal invariance. It is important that $\f$ and $s$ be isolated in the operator spectrum for us to be able to exploit this constraint. For instance, imagine there were two singlet scalars $s_{1,2}$ with the same dimension. Then $(\l_{\f\f s}^\mathrm{fake})^2=\l_{\f\f s_1}^2+\l_{\f\f s_2}^2$ would appear in (\ref{eq:vectoreq}). This combination does not satisfy $\l_{\f\f s}^\mathrm{fake}=\l_{\f s_i\f}$.

\item We will sometimes assume additional gaps to derive lower bounds on OPE coefficients.  For instance, to obtain a lower bound on the coefficient of the conserved $O(N)$ current in the $\f_i\x\f_j$ OPE, we will need to assume a gap between the first and second spin-1 antisymmetric tensor operators.
\end{itemize}

As an example, (\ref{eq:example}) shows a semidefinite program that incorporates symmetry of $\l_{\f\f s}$ and the assumption that $\f_i,s$ are the only relevant scalars in their respective sectors:
\be
\label{eq:example}
\begin{array}{cll}
\left(\begin{array}{ccc} 1 & 1 \end{array}\right) \vec\alpha\cdot \vec{V}_{S,0,0} \left( \begin{array}{c} 1 \\ 1 \end{array} \right) \geq 0,&\textrm{(unit operator)}&\\
\vec\alpha\cdot \vec{V}_{T,\Delta,\ell} \geq 0,&\Delta\geq\frac{D-2}2, & \ell=0,\textrm{ and} \\ 
 &\Delta\geq\ell+D-2, &\ell>0 \textrm{ even}; \\
\vec\alpha\cdot \vec{V}_{A,\De,\ell}  \geq 0,& \Delta\geq\ell+D-2, &\ell\textrm{ odd};\\
\vec\alpha\cdot \vec{V}_{V,\De,\ell} \geq 0,&\Delta\geq D &\ell=0, \textrm{ and}\\
& \Delta\geq\ell+D-2, &\ell>0;\\
\vec\alpha\cdot \vec{V}_{S,\De,\ell} \succeq 0,& \Delta\geq D, &\ell=0,\textrm{ and}\\
& \Delta\geq\ell+D-2 & \ell>0\textrm{ even};\\
\vec\alpha\cdot \left(\vec{V}_{S,\Delta_s,0} + \vec{V}_{V,\Delta_\phi,0} \otimes \left(\begin{array}{cc}1 &0 \\0 &0\end{array}\right)\right)   \succeq 0 .
\end{array}
\ee
The final constraint in (\ref{eq:example}) imposes the appearance of $\f_i,s$ in the OPEs and incorporates the equality $\l_{\f\f s}=\l_{\f s\f}$.\footnote{In writing this constraint, we have assumed the scalar conformal blocks are normalized so that $g_{\Delta,\ell}(u,v) \sim C u^{\Delta/2}$ to leading order in $u$, where $C$ is a $\Delta$-independent constant.} It replaces two otherwise independent constraints on $V_{S}$ and $V_{V}$.  As previously mentioned, if we assume no gap between $\phi_i$, $s$ and the next operators in each sector, enforcing symmetry of the OPE coefficients will have no effect: indeed each of the terms in this constraint would be independently positive-semidefinite, since the other inequalities imply $\vec\alpha\cdot\vec{V}_{S,\Delta_ s+\delta,0}\succeq0$ and $\vec\alpha\cdot \vec{V}_{V,\Delta_{\phi}+\delta,0} \geq 0$ for $\delta$ arbitrary small. 

Finally, one might want to enforce the existence of a unique relevant scalar operator, with dimension $\Delta_t$, transforming in the traceless symmetric representation. In this case the symmetric tensor constraint is replaced by
\begin{align}
\vec\alpha\cdot \vec{V}_{T,\Delta,\ell} \geq 0,\qquad\qquad& \Delta=\Delta_t \text{ or } \Delta>D, \qquad \ell=0,\text{ and}\nonumber\\ 
&  \Delta\geq\ell+D-2, \qquad\,\,\,\,\,\,\,\, \ell>0 \text{ even}.\label{eq:functional-tensor} 
\end{align}

\section{Results}
\label{sec:results}

\subsection{$O(2)$}

\begin{figure}[htbp]
\begin{center}
\includegraphics[scale=1]{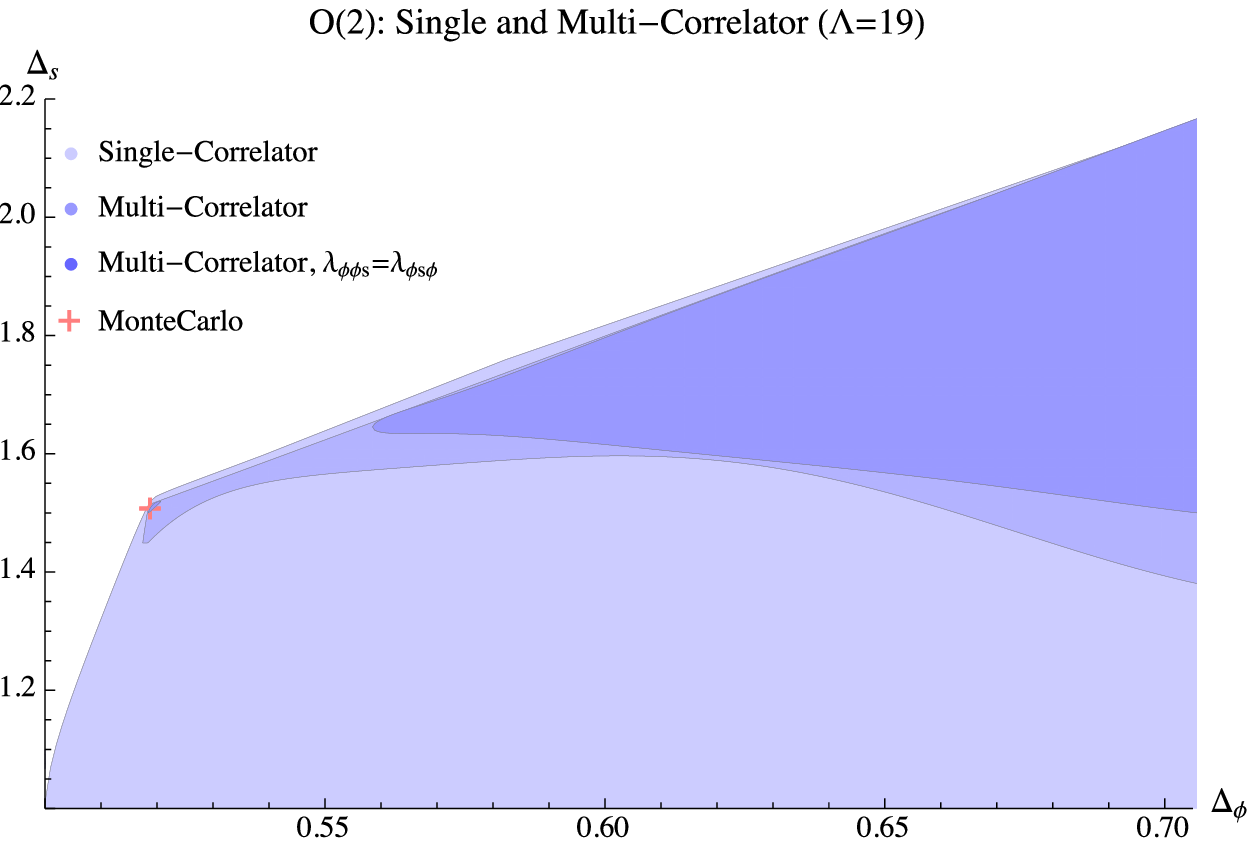}
\caption{Allowed region for $(\Delta_\phi,\Delta_s)$ in 3d CFTs with $O(2)$ symmetry. The light blue region makes no additional assumptions and was computed in \cite{Kos:2013tga} using the correlator $\<\f\f\f\f\>$ at $\Lambda=19$. The medium blue region was computed from the system of correlators $\<\f\f\f\f\>$, $\<\f\f s s\>$, $\<s s s s\>$ at $\Lambda=19$, and assumes $\Delta_\f$ and $\Delta_s$ are the only relevant dimensions in the vector and singlet scalar channels at which contributions appear. The dark blue region is computed similarly, but additionally assumes the OPE coefficient relation $\l_{\f\f s} = \l_{\f s \f}$. This latter assumption leads to a small closed region in the vicinity of the red cross, which represents the Monte Carlo estimate for the position of the $O(2)$ model from \cite{Campostrini:2006ms}.} 
\label{fig:O2singlecorrelator}
\end{center}
\end{figure}

To begin, let us recall the bounds on $\Delta_\f, \Delta_s$ computed in \cite{Kos:2013tga} using the correlation function $\<\f_i\f_j\f_k\f_l\>$ (see figure~\ref{fig:O2singlecorrelator}).  Like the Ising model bounds computed in \cite{ElShowk:2012ht,El-Showk:2014dwa}, this single-correlator bound has an excluded upper region, an allowed lower region, and a kink in the curve separating the two.  The position of this kink corresponds closely to where we expect the $O(2)$ model to lie, and one of our goals is to prove using the bootstrap that the $O(2)$ model does indeed live at the kink.\footnote{The sharpness of the kink depends on the number of derivatives $\Lambda$ used when computing the bound (appendix \ref{sec:appA}). Figure~\ref{fig:O2singlecorrelator} was computed at a lower derivative order than we use for most of this work, so the kink is relatively smooth.}  If we assume that $s$ is the only relevant $O(2)$ singlet, then a small portion of the allowed region below the kink gets carved away, analogous to the Ising case in \cite{Kos:2014bka}.

Adding the constraints of crossing symmetry and unitarity for the full system of correlators $\<\f\f\f\f\>,\<\f\f s s\>,\<s s s s\>$ does not change these bounds without additional assumptions.  However, having access to the correlator $\<\f\f s s\>$ lets us input information special to the $O(2)$ model that does have an effect.  We expect that $\f$ is the only relevant $O(2)$ vector in the theory.  One way to understand this fact is via the equation of motion at the Wilson-Fisher fixed point in $4-\epsilon$ dimensions,
\be
\square \phi_i &\propto & \lambda \phi^2 \phi_i.
\ee
This equation implies that the operator $\phi^2 \phi_i$ is a descendent, so there is a gap in the spectrum of $O(2)$-vector primaries between $\phi_i$ and the next operator in this sector, which is a linear combination of $\phi_i \phi^4$ and $\phi_i (\partial\phi)^2$.  The equation of motion makes sense in perturbation theory $\e\ll 1$.  However, it is reasonable to expect gaps in the spectrum to be robust as $\e$ gets larger.  In particular, we expect this gap to persist as $\e\to 1$.  Thus, a gap in the $O(2)$-vector sector reflects the equations of motion of the $O(2)$ model.

We do not know if there is sharp experimental evidence for the claim that the $O(2)$ model contains exactly one relevant $O(2)$-vector scalar.  The cleanest experimental realization of the $O(2)$ model is the superfluid transition in ${}^4$He~\cite{Lipa:2003zz}.  This theory has microscopic $O(2)$ symmetry, so one cannot easily determine the number of relevant $O(2)$-vector scalars by counting order parameters.  The number could be determined by counting order parameters in systems where the $O(2)$ symmetry is emergent. 

As explained above, it is natural to impose a gap in both the $O(2)$ vector and singlet sectors in our formalism, giving rise to the medium blue region in figure~\ref{fig:O2singlecorrelator}. Another important constraint is symmetry of the OPE coefficient $\l_{\f\f s} = \l_{\f s \f}$.  Adding this constraint gives the dark blue region in figure~\ref{fig:O2singlecorrelator}; a close-up view of the $O(2)$ model point is shown in figure~\ref{fig:O2island}, which we show for increasing numbers of derivatives $\Lambda=19,27,35$ (see appendix~\ref{sec:appA}). We now have a closed island around the expected position of the $O(2)$ model, very close to the original kink in figure~\ref{fig:O2singlecorrelator}. The bounds strengthen as $\Lambda$ increases.  However, the allowed regions apparently do not shrink as quickly as in the case of the 3d Ising CFT \cite{Simmons-Duffin:2015qma}.  Thus, our determination of $(\Delta_\phi,\Delta_s)$ is unfortunately not competitive with the best available Monte Carlo~\cite{Campostrini:2006ms} and experimental~\cite{Lipa:2003zz} results (though it is consistent with both).\footnote{Note that ${}^4$He experiments cannot easily determine $\De_\phi$ because the $O(2)$ symmetry is realized microscopically. Some results constraining $\De_\phi$ have been reported from NMR experiments (e.g., as summarized in~\cite{Pelissetto:2000ek}) but they are not very precise.}  
We conjecture that including additional crossing relations (such as those involving the symmetric tensor $t_{ij}$) will give even stronger bounds; we plan to explore this possibility in future work.

\begin{figure}[tbp]
\begin{center}
\includegraphics[scale=1]{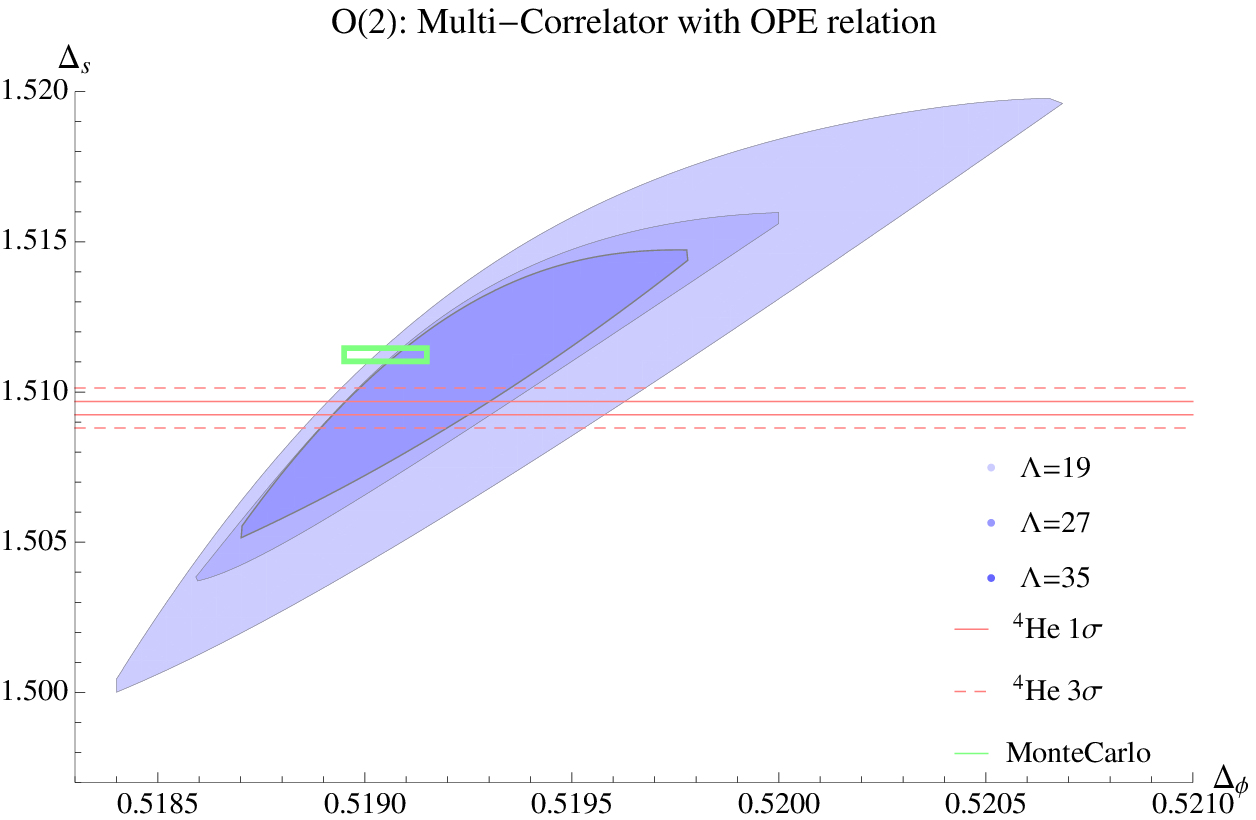}
\caption{Allowed regions for $(\Delta_\phi,\Delta_s)$ in 3d CFTs with $O(2)$ symmetry and exactly one relevant $O(2)$ vector $\phi$ and singlet $s$, computed from the system of correlators $\<\f\f\f\f\>$, $\<\f\f s s\>$, and $\< ssss\>$ using \texttt{SDPB} with $\Lambda=19,27$, and $35$ (see appendix~\ref{sec:appA}).  The smallest region (darkest blue) corresponds to $\Lambda=35$.  The green rectangle represents the Monte Carlo estimate \cite{Campostrini:2006ms}.  The red lines represent the 1$\s$ (solid) and 3$\s$ (dashed) confidence intervals for $\Delta_s$ from experiment \cite{Lipa:2003zz}.  The allowed/disallowed regions in this work were computed by scanning over a lattice of points in operator dimension space.  For visual simplicity, we fit the boundaries with curves and show the resulting curves.  Consequently, the actual position of the boundary between allowed and disallowed is subject to some error (small compared to size of the regions themselves). We tabulate this error in appendix~\ref{sec:appA}.}
\label{fig:O2island}
\end{center}
\end{figure}

\begin{figure}[htbp]
\begin{center}
\includegraphics[scale=0.8]{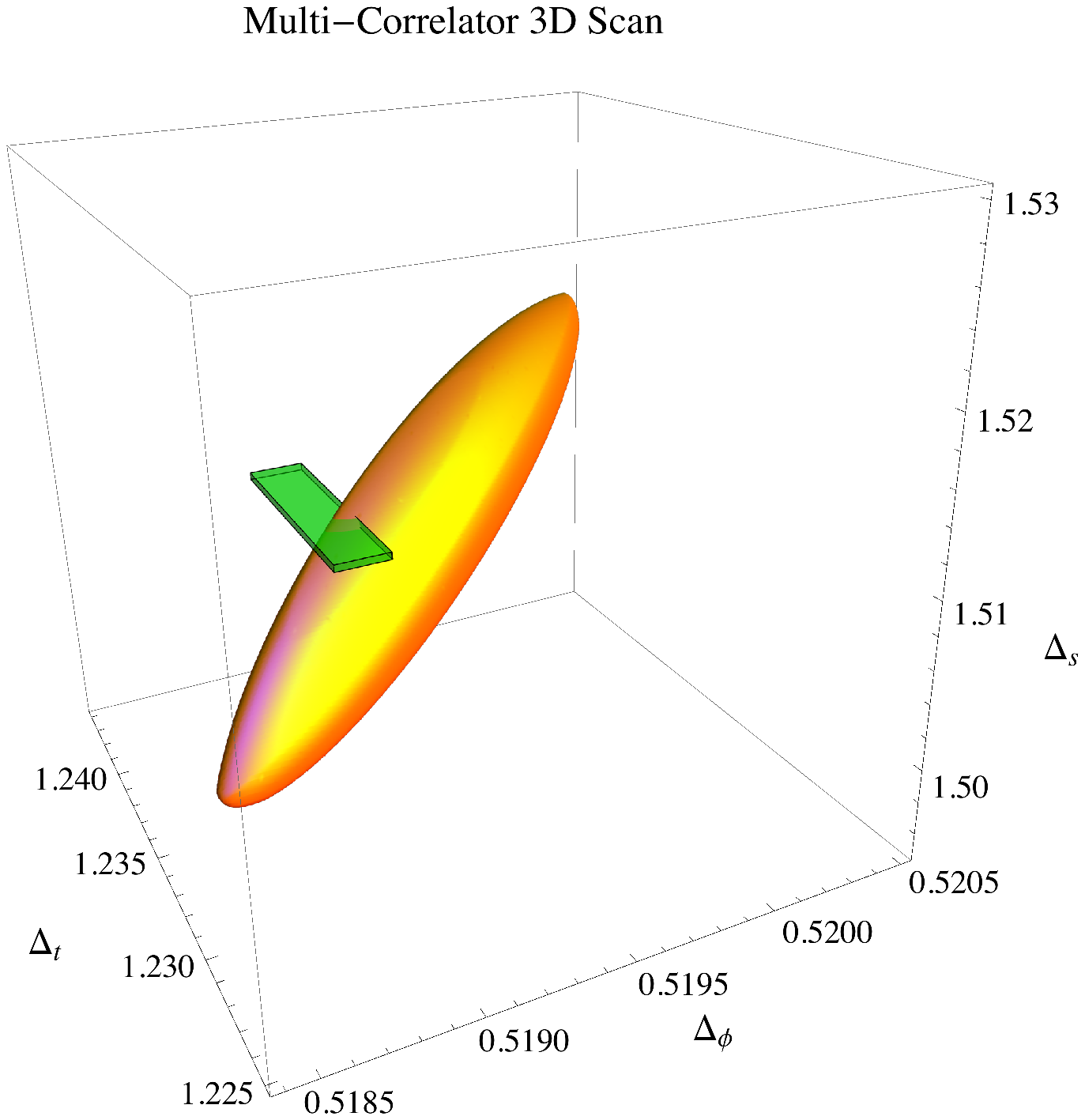}
\caption{Allowed region (orange) for $(\Delta_\phi,\Delta_s,\Delta_t)$ in a 3d CFT with $O(2)$ symmetry and exactly one relevant $O(2)$-vector $\phi$, $O(2)$ singlet $s$, and $O(2)$ traceless symmetric-tensor $t$. This region was computed using \texttt{SDPB} with $\Lambda=19$. The green rectangle represents the error bars from Monte Carlo \cite{Campostrini:2006ms} and the pseudo-$\epsilon$ expansion approach \cite{Calabrese:2004ca}. Note that our estimate for $\Delta_t$ in (\ref{eq:symtensorestimate}) was computed with $\Lambda=35$, so it is more precise than the region pictured here.
}
\label{fig:O2island3d}
\end{center}
\end{figure}

In addition to gaps in the $O(2)$-vector and singlet sectors, we also expect that the $O(2)$ model has a single relevant traceless symmetric tensor $t_{ij}$.  Let us finally impose this condition by demanding that $t_{ij}'$ has dimension above $D=3$ and scanning over $\Delta_t$ along with $\Delta_\phi, \Delta_s$.  The result is a three-dimensional island for the relevant scalar operator dimensions, which we show in figure~\ref{fig:O2island3d}.  Our errors for the symmetric-tensor dimension $\Delta_t$ are much more competitive with previous determinations.  By scanning over different values of $(\Delta_\phi,\Delta_s)$ in the allowed region and computing the allowed range of $\Delta_t$ at $\Lambda=35$, we estimate
\be
\label{eq:symtensorestimate}
1.2325 < \Delta_t < 1.239 \qquad\qquad\textrm{($O(2)$ model)}\,,
\ee
which is consistent with previous results from the pseudo-$\epsilon$ expansion approach~\cite{Calabrese:2004ca} giving $\Delta_t = 1.237(4)$.

\subsection{$O(N)$, $N>2$}

\begin{figure}[htbp]
\begin{center}
\includegraphics[scale=1]{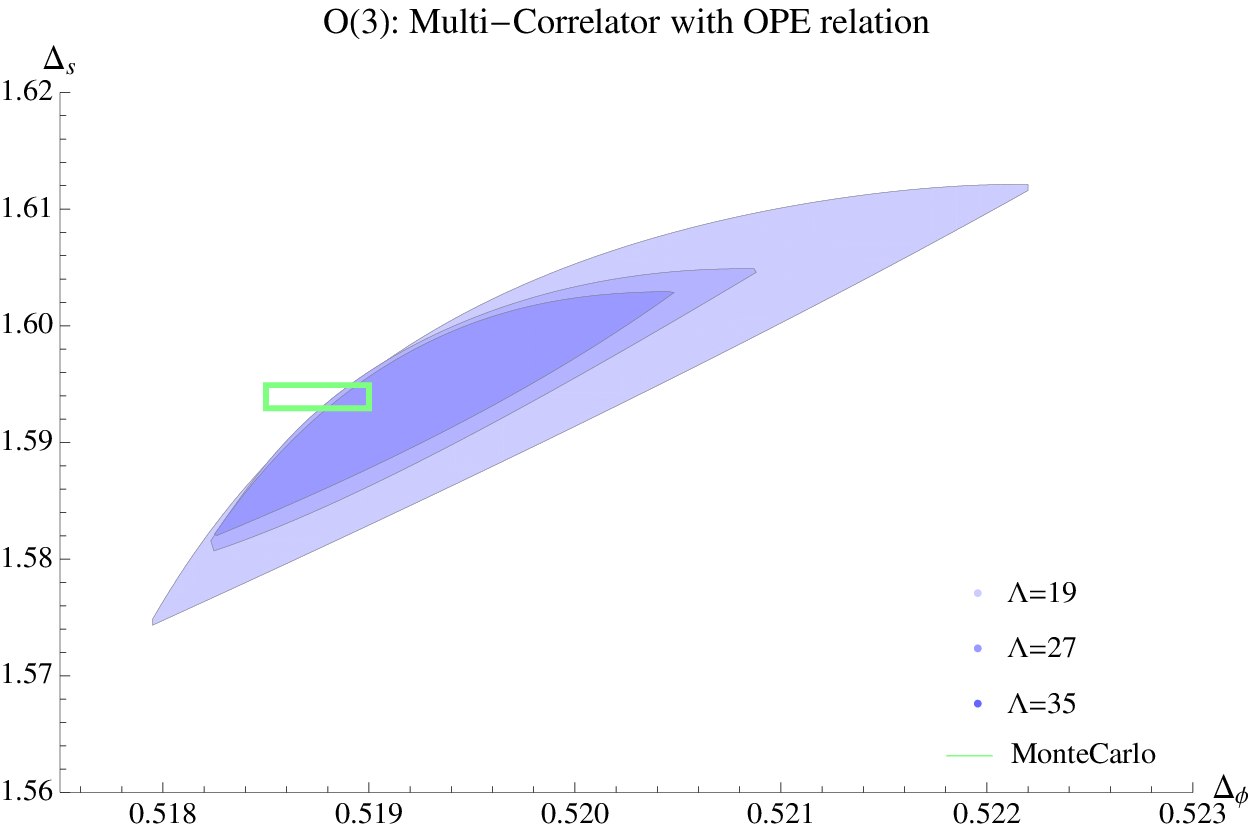}
\caption{Allowed regions for $(\Delta_\phi,\Delta_s)$ in 3d CFTs with $O(3)$ symmetry and exactly one relevant $O(3)$-vector $\phi$ and $O(3)$ singlet $s$, computed using \texttt{SDPB} with $\Lambda=19,27$, and $35$ (see appendix~\ref{sec:appA}).  The smallest region (darkest blue) corresponds to $\Lambda=35$.  The green rectangle represents the Monte Carlo estimate \cite{Campostrini:2002ky}.}
\label{fig:O3island}
\end{center}
\end{figure}

The bounds for $N>2$ are similar to the case of $N=2$.  In figure~\ref{fig:O3island}, we show the allowed region of $(\Delta_\phi,\Delta_s)$ for theories with $O(3)$ symmetry, assuming $\phi$ and $s$ are the only relevant scalars in their respective $O(N)$ representations, and using symmetry of the OPE coefficient $\l_{\f\f s}$.  We expect that an additional scan over $\Delta_t$ would yield a 3d island similar to figure~\ref{fig:O2island3d}.  By performing this scan at a few values of $(\Delta_\phi,\Delta_s)$, we estimate
\be
1.204 < \Delta_t < 1.215 \qquad\qquad \textrm{($O(3)$ model)}\,,
\ee
which is consistent with previous results from the pseudo-$\epsilon$ expansion approach~\cite{Calabrese:2004ca} giving $\Delta_t = 1.211(3)$.

\begin{figure}[htbp]
\begin{center}
\includegraphics[scale=1]{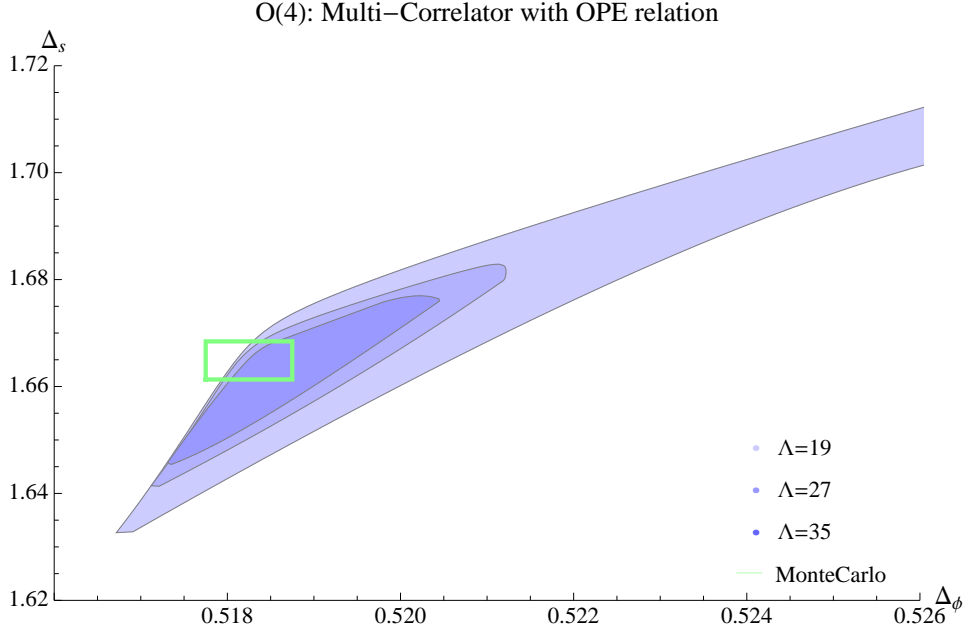}
\caption{Allowed regions for $(\Delta_\phi,\Delta_s)$ in 3d CFTs with $O(4)$ symmetry and exactly one relevant $O(4)$-vector $\phi$ and $O(4)$ singlet $s$, computed using \texttt{SDPB} with $\Lambda=19,27$, and $35$ (see appendix~\ref{sec:appA}).  The smallest region (darkest blue) corresponds to $\Lambda=35$.  The green rectangle represents the Monte Carlo estimate \cite{Hasenbusch:2000ph}.}
\label{fig:O4island}
\end{center}
\end{figure}

In figure~\ref{fig:O4island}, we show the allowed region of $(\Delta_\phi,\Delta_s)$ for the $O(4)$ model, with the same assumptions as discussed above for $O(3)$.  A clear trend is that the allowed region is growing with $N$.  For example, at $\Lambda=19$, the $O(4)$ allowed region isn't even an island --- it connects to a larger region not shown in the plot.  Increasing the number of derivatives to $\Lambda=35$ shrinks the region, but it is not as small as in the case of $O(2)$ or $O(3)$.  

The trend of lower-precision determinations at larger $N$ reverses at some point.  For example, in figure~\ref{fig:Archipelago}, the allowed region for $N=20$ is smaller again than the $O(4)$ region. The relative size of the $O(4)$ region and the $O(20)$ region is $\Lambda$-dependent, and we have not studied the pattern for general $N$ in detail. However, as an important check we note that the $O(20)$ island in figure~\ref{fig:Archipelago} is nicely compatible with the $1/N$ expansion (see~\cite{Kos:2013tga}), giving the point $(\Delta_{\f}, \Delta_s) \simeq (.5064,1.938)$ which sits in the upper-left corner of the allowed region.

Finally, we remark that all of the constraints on operator dimensions found above can be reinterpreted in terms of constraints on critical exponents. Following standard critical exponent notation (see~\cite{Pelissetto:2000ek}), the relations are given by
\begin{align}
 &\eta = 2 \De_{\phi} -1, \qquad\nu = \frac{1}{3-\De_{s}}, \qquad\gamma = \frac{3-2\De_{\phi}}{3-\De_{s}}, \qquad\alpha = \frac{3-2\De_{s}}{3-\De_{s}}, \nn\\
 &\beta = \frac{\De_{\phi}}{3-\De_{s}}, \qquad\,\,\delta = \frac{3-\De_{\phi}}{\De_{\phi}}, \qquad\zeta = \frac{3-4\De_{\phi}}{3-\De_{s}}, \qquad\phi_2 = \frac{3-\De_t}{3-\De_s}\,.
\end{align}

\subsection{Current Central Charges}
Let $J_{ij}^\mu(x)$ be the conserved currents that generate $O(N)$ transformations. $J_{ij}^\mu(x)$ is an $O(N)$ antisymmetric tensor with spin 1 and dimension 2. Its 2-point function is determined by conformal and $O(N)$ symmetry to be
\be
\label{eq:jjtwopoint}
\< J^{\mu}_{ij}(x_1) J^\nu_{kl}(x_2) \> =(\delta_{ik}\delta_{jl} - \delta_{il} \delta_{jk})  \frac{C_J}{(4\pi)^2} \frac{1}{x_{12}^4}\left[ \eta^{\mu\nu} -2 \frac{(x_1-x_2)^\mu (x_1-x_2)^\nu}{x_{12}^2} \right].
\ee
We call the normalization coefficient $C_J$ from Eq.~(\ref{eq:jjtwopoint}) the current central charge.\footnote{This name is by analogy with the case of 2d CFTs, where $C_J$ appears as a central element in an affine Kac-Moody algebra. In higher dimensional CFTs, $C_J$ is not an element of a nontrivial algebra in general, though it can be in special cases \cite{Beem:2013sza}.} The conserved current $J^\mu_{ij}$ appears in the sum over antisymmetric-tensor operators $\mathcal{O}_A$ in Eq.~(\ref{eq:vectoreq}). A Ward identity relates the OPE coefficient $\lambda_{J\phi\phi}$ to $C_J$. In our conventions
\be
\lambda_{\phi\phi J}^2 = \frac{8}{C_J/C_J^{\text{free}}}\,,
\ee
where $C_J^{\text{free}} = 2$ is the free theory value of $C_J$ \cite{Osborn:1993cr, Petkou:1994ad}. In the $O(N)$ vector models $C_J$ is known to have the large $N$ and $\epsilon$ expansions \cite{Petkou:1995vu}
\be
\frac{C_J}{C_J^{\text{free}}} \bigg|_{d=3} = 1- \frac{32}{9\pi^2} \frac{1}{N} + O\left(\frac{1}{N^2}\right),\qquad \frac{C_J}{C_J^{\text{free}}}\bigg|_{d=4-\epsilon} = 1 - \frac{3(N+2)}{4(N+8)^2} \epsilon^2 + O(\epsilon^3).
\ee
Note that both of these expansions predict that $C_J$ will be smaller than the free value.

It is well known that the conformal bootstrap allows one to place upper bounds on OPE coefficients, or equivalently a lower bound on $C_J$. Previously such bounds were explored in $d=4$ in \cite{Poland:2010wg,Poland:2011ey} and in $d=3,5$ in \cite{Nakayama:2014yia}. To find such a bound, we search for a functional $\alpha$ with the following properties (cf. eq.~(\ref{eq:example})):
\be
\label{eq:cjsdp}
\begin{array}{cll}
\vec\alpha\cdot \vec{V}_{A,2,1} = 1,&\textrm{(normalization)}&\\
\vec\alpha\cdot \vec{V}_{T,\Delta,\ell} \geq 0,&\Delta\geq\frac{D-2}2, & \ell=0,\textrm{ and} \\ 
 &\Delta\geq\ell+D-2, &\ell>0 \textrm{ even}; \\
\vec\alpha\cdot \vec{V}_{A,\De,\ell}  \geq 0,& \Delta\geq\ell+D-2, &\ell\textrm{ odd};\\
\vec\alpha\cdot \vec{V}_{V,\De,\ell} \geq 0,&\Delta\geq D&\ell=0, \textrm{ and}\\
& \Delta\geq\ell+D-2, &\ell>0;\\
\vec\alpha\cdot \vec{V}_{S,\De,\ell} \succeq 0,& \Delta\geq D, &\ell=0,\textrm{ and}\\
& \Delta\geq\ell+D-2 & \ell>0\textrm{ even};\\
\vec\alpha\cdot \left(\vec{V}_{S,\Delta_s,0} + \vec{V}_{V,\Delta_\phi,0} \otimes \left(\begin{array}{cc}1 &0 \\0 &0\end{array}\right)\right)   \succeq 0 \,.
\end{array}
\ee
Notice that compared to (\ref{eq:example}), we have dropped the assumption of the functional $\vec \alpha$ being positive on the identity operator contribution and we chose a convenient normalization for $\vec \alpha$. It follows then from the crossing equation (\ref{eq:vectoreq}) that
\be
\label{eq:cjbound}
\frac{8}{C_J/C_J^{\text{free}}} \le - \left(\begin{array}{ccc} 1 & 1 \end{array}\right) \vec\alpha\cdot \vec{V}_{S,0,0} \left( \begin{array}{c} 1 \\ 1 \end{array} \right)\,.
\ee
Therefore, finding a functional $\vec \alpha$ sets a lower bound on $C_J$. To improve the bound, we should minimize the RHS of (\ref{eq:cjbound}). We thus seek to minimize
\be
- \left(\begin{array}{ccc} 1 & 1 \end{array}\right) \vec\alpha\cdot \vec{V}_{S,0,0} \left( \begin{array}{c} 1 \\ 1 \end{array} \right)\,,
\ee
subject to the constraints (\ref{eq:cjsdp}). This type of problem can be efficiently solved using \texttt{SDPB}. In this way, we set a lower bound on $C_J$ for all allowed values of $\De_\f$, $\De_s$.

We can also set an upper bound on $C_J$, provided we additionally assume a gap in the spin-1 antisymmetric tensor sector. At this point it is not clear what gap we should assume, but to stay in the spirit of our previous assumptions, we will assume that the dimension of the second spin-1 antisymmetric tensor satisfies $\Delta_{J'} \geq 3$, so that the current $J_{ij}^\mu$ is the only relevant operator in this sector.  We now search for a functional $\vec\alpha$ (different from the one above) that satisfies
\be
\vec\alpha\cdot \vec{V}_{A,\De,1}  \geq 0,& \Delta\geq 3, &\\
\vec\alpha\cdot \vec{V}_{A,\De,\ell}  \geq 0,& \Delta\geq\ell+D-2, &\ell>1 \textrm{ odd},
\ee
and is normalized so that
\be
\vec\alpha\cdot \vec{V}_{A,2,1} = -1\,.
\ee
The constraints on $\vec\alpha$ coming from the singlet and traceless symmetric-tensor sectors stay the same as in (\ref{eq:cjsdp}). An upper bound on $C_J$ then follows from (\ref{eq:vectoreq}):
\be
\label{eq:cjupperbound}
\frac{8}{C_J/C_J^{\text{free}}} \ge  \left(\begin{array}{ccc} 1 & 1 \end{array}\right) \vec\alpha\cdot \vec{V}_{S,0,0} \left( \begin{array}{c} 1 \\ 1 \end{array} \right)\,.
\ee

\begin{figure}
\begin{center}
\begin{tabular}{c c}
\raisebox{-.5\height}{\includegraphics[scale=0.55]{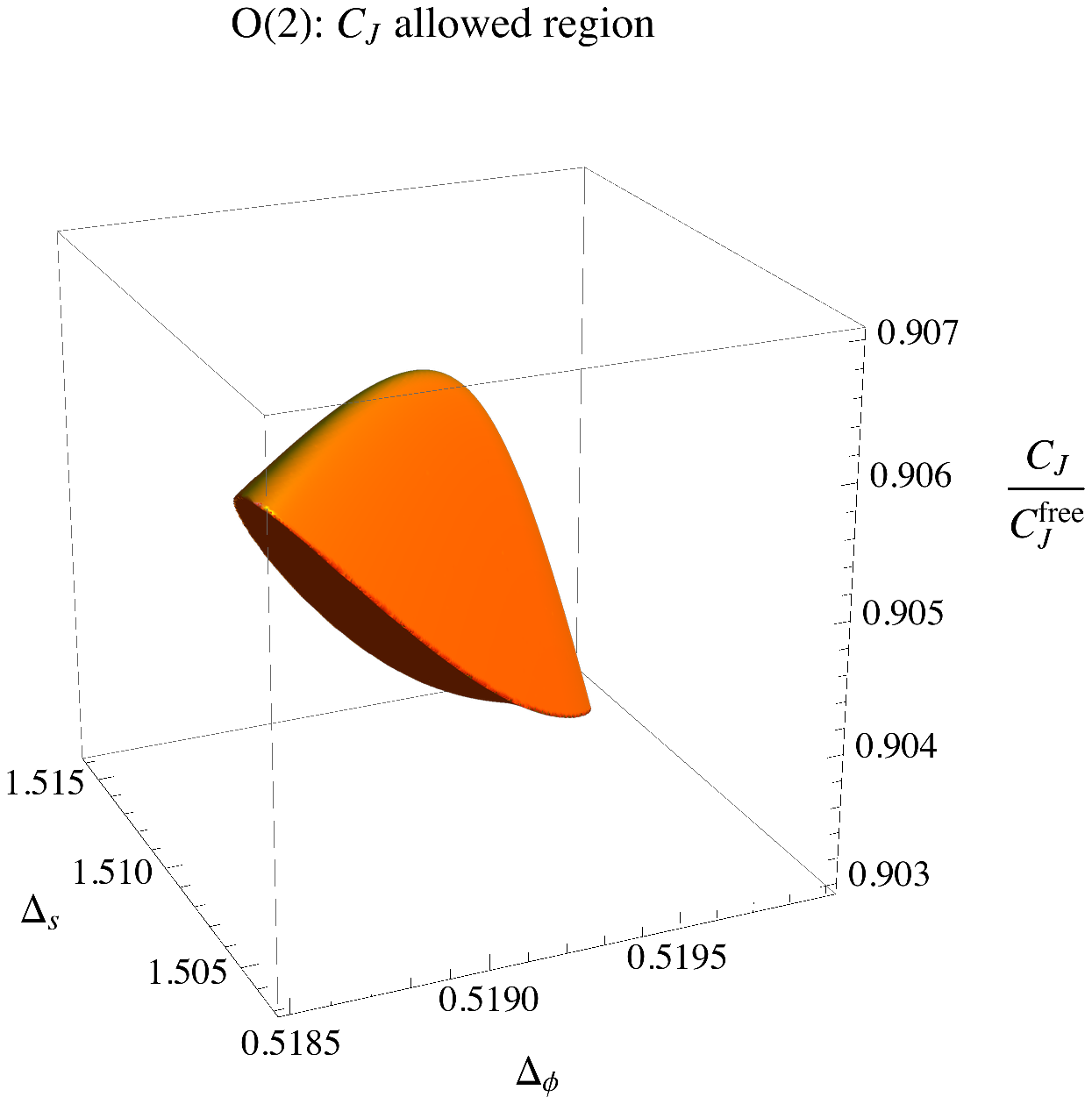}} &\raisebox{-.5\height}{ \includegraphics[scale=0.6]{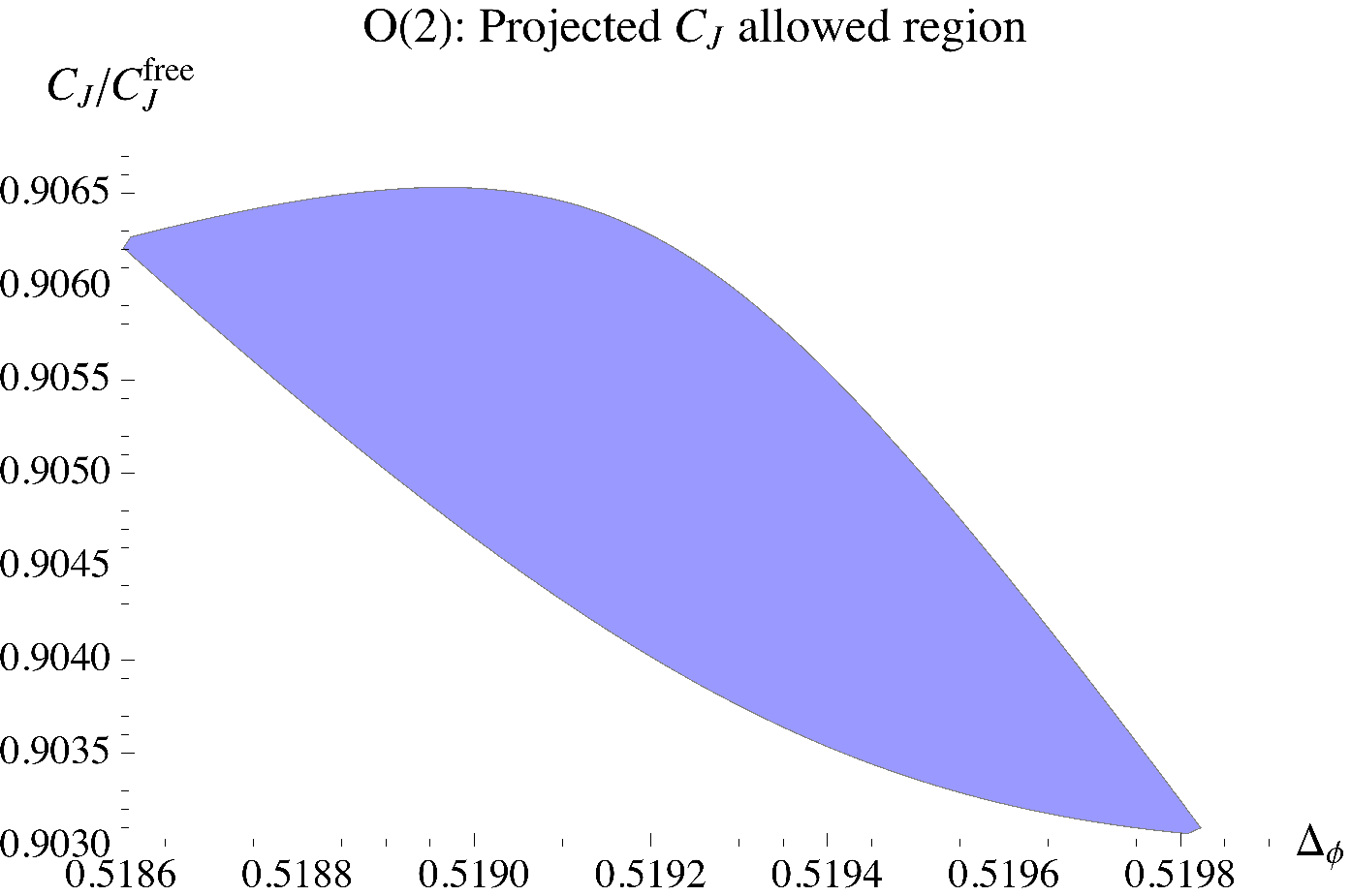} }
\end{tabular}
\caption{The left panel shows the allowed values of $C_J$ as a function of $\De_\f$ and $\De_s$ in $O(2)$ symmetric theories. The right panel is the projection of the allowed region onto the $(\De_\f, C_J)$ plane. Both plots are computed using \texttt{SDPB} with $\Lambda=27$.}\label{fig:O2CJ}
\end{center}
\end{figure}

\begin{figure}
\begin{center}
\begin{tabular}{c c}
\raisebox{-.5\height}{\includegraphics[scale=0.55]{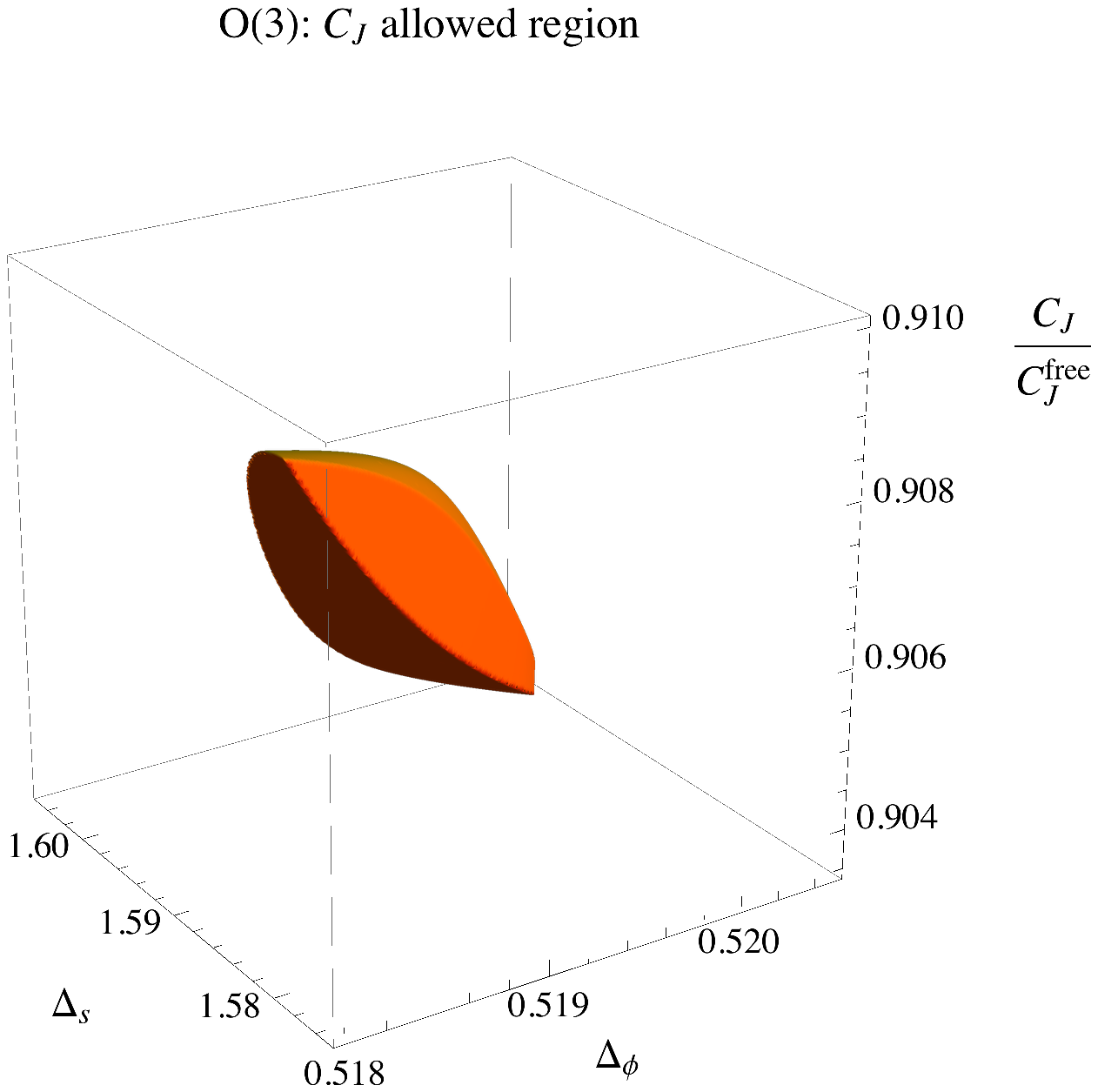}} &\raisebox{-.5\height}{ \includegraphics[scale=0.6]{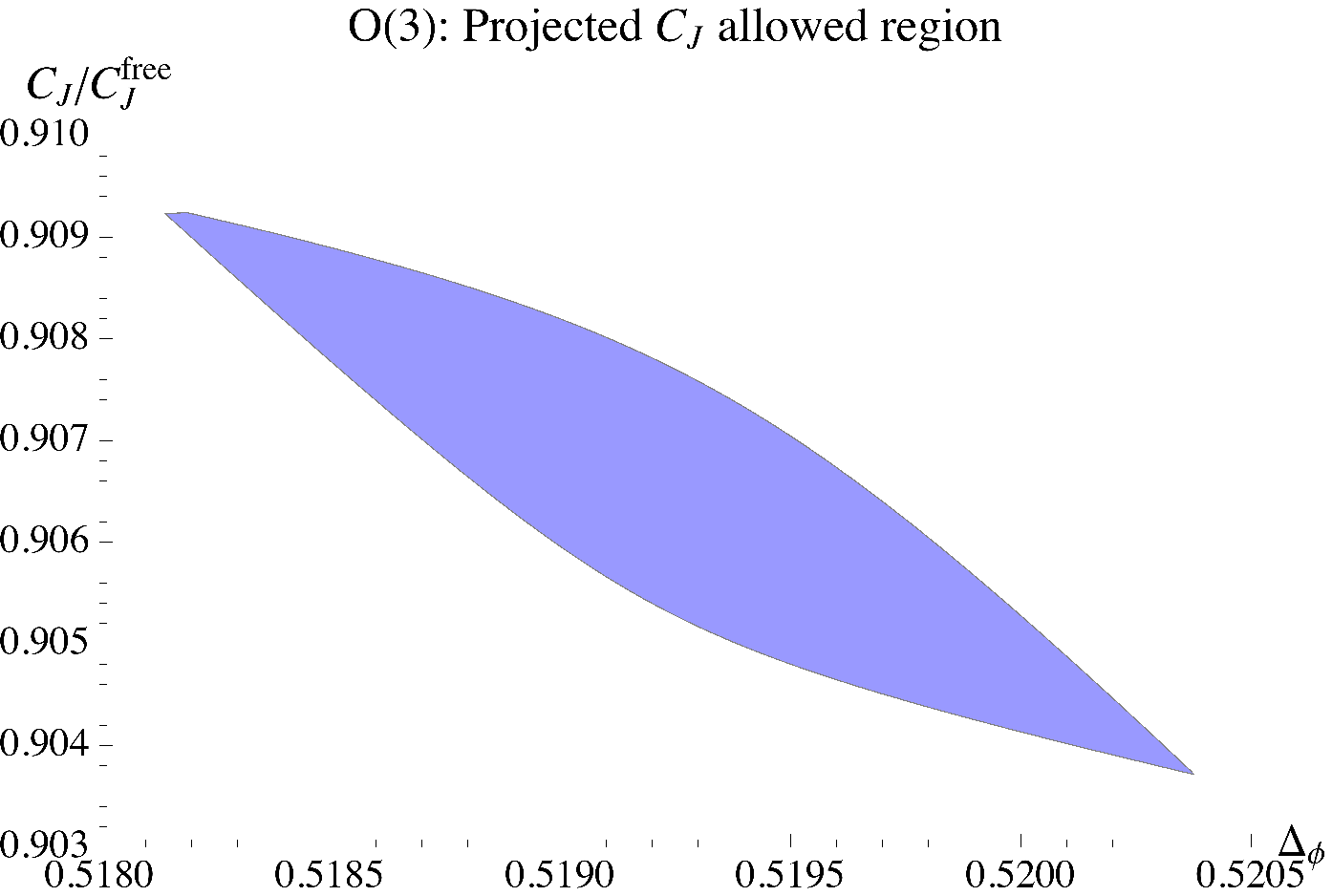} }
\end{tabular}
\caption{The left panel shows the allowed values of $C_J$ as a function of $\De_\f$ and $\De_s$ in $O(3)$ symmetric theories. The right panel is the projection of the allowed region onto the $(\De_\f, C_J)$ plane. Both plots are computed using \texttt{SDPB} with $\Lambda=27$.}\label{fig:O3CJ}
\end{center}
\end{figure}

Our upper and lower bounds on $C_J$, expressed as a function of $\Delta_\phi$ and $\Delta_s$, are shown in figures \ref{fig:O2CJ} and \ref{fig:O3CJ} for $O(2)$ and $O(3)$ symmetry, respectively. The allowed region for a given $N$ consists of a 3d island in $(\Delta_\phi, \Delta_s, C_J)$ space. This determines the current central charge to within the height of the island. For the two physically most interesting cases, $N=2$ and $N=3$, we find:
\be
N=2:\quad \frac{C_J}{C_J^{\text{free}}} = 0.9050(16)\,, && N=3:\quad \frac{C_J}{C_J^{\text{free}}} =0.9065(27) \,.
\ee
As an additional check, we also computed $C_J$ for $N=20$:
\be
N=20:\quad \frac{C_J}{C_J^{\text{free}}} = 0.9674(8) \,.
\ee
This result agrees within $0.5\%$ accuracy with the leading $1/N$ expansion result, $C_J/C_J^{\text{free}} \approx 0.964$ \cite{Petkou:1995vu}.

Recently, the current central charge attracted some interest in studies of transport properties of $O(N)$ symmetric systems near a quantum critical point, where $C_J$ can be related to the conductivity at zero temperature. In particular, using the OPE it was found in \cite{Katz:2014rla} that the asymptotic behavior of conductivity at low temperature is given by
\be
\label{eq:conductivity}
\frac{\sigma(\omega/T)}{\sigma_Q} = \sigma_\infty + B \mathcal{C} i^{\Delta_s} \left( \frac{T}{\omega} \right)^{\Delta_s} - i 24 C_T \gamma H_{xx} \left( \frac{T}{\omega}\right)^3 + \dots\,,
\ee
where $\sigma_Q = e^2/\hbar$ is the conductance quantum. Here, $\sigma_\infty$ is the (unitless) conductivity at high frequency and zero temperature which is related to $C_J$ as
\be
\sigma_\infty = C_J/32\,.
\ee
Furthermore, $C_T$ is the central charge of the theory, $\mathcal C$ is the $\<JJs\>$ OPE coefficient, and $\gamma$ is one of the $\<JJT\>$ OPE coefficients, where $T$ is the energy-momentum tensor. $B$ and $H_{xx}$ are the finite temperature one-point function coefficients:
\be
\< s \>_T = B T^{\Delta_s}\,, \qquad \< T_{xx} \>_T = H_{xx} T^3\,.
\ee
Of all the CFT data that goes into (\ref{eq:conductivity}), we have determined $\sigma_\infty$ and $\Delta_s$ for the $O(N)$ vector models in this work, while $C_T$ was estimated using bootstrap methods before in \cite{Kos:2013tga}. The OPE coefficients $\mathcal C$ and $\gamma$ can not be determined in our setup, but could in principle be obtained by including the conserved current $J^\mu_{ij}$ as an external operator in the crossing equations. The one-point functions $B$ and $H_{xx}$ are in principle determined by the spectrum and OPE coefficients of the theory \cite{ElShowk:2011ag}. However, to compute them we would need to know the high-dimension operator spectrum. This is still out of the reach of the conformal bootstrap approach. 

Of particular interest for physical applications is the $N=2$ case, which describes superfluid-insulator transitions in systems with two spatial dimensions \cite{PhysRevB.44.6883,PhysRevLett.95.180603}. Some examples of such systems are thin films of superconducting materials, Josephson junction arrays, and cold atoms trapped in an optical lattice. In these systems the parameter $\sigma_\infty$ is the high-frequency limit of the conductivity. This quantity has not yet been measured in experiments, but was recently computed in Quantum Monte Carlo simulations \cite{Witczak-Krempa:2013nua,Katz:2014rla}, \cite{Chen:2014}, and \cite{Gazit:2014} to be $2\pi\sigma_\infty^{\mathrm{MC}}= 0.3605(3), 0.359(4)$, and $0.355(5)$, respectively\footnote{These uncertainties reflect statistical errors but may not include systematic effects, conservatively estimated in \cite{Witczak-Krempa:2013nua} to be $5-10\%$. We thank Subir Sachdev, Erik S{\o}rensen, and William Witczak-Krempa for correspondence on this point.}. Our rigorous result $2\pi\sigma_\infty^{\mathrm{Bootstrap}} = 0.3554(6)$ is in excellent agreement with these determinations and is significantly more precise after systematic uncertainties are taken into account.

\section{Conclusions}
\label{sec:conclusions}

In this work, we used the conformal bootstrap with multiple correlators to set more stringent bounds on the operator spectrum of 3d CFTs with $O(N)$ symmetry. The multiple correlator approach works in this setting similarly to the case of $\mathbb{Z}_2$-symmetric CFTs -- including mixed correlators opens access to parts of the spectrum that are inaccessible with a single correlator. In this work we considered mixed correlators of an $O(N)$ singlet and an $O(N)$ vector, gaining access to the sector of $O(N)$ vectors. We can then additionally input assumptions about the operator spectrum in that sector. As a result, we exclude large portions of the allowed space of CFTs. This reaffirms conclusions from previous works on the 3d Ising model: it is important and fruitful to consider multiple crossing equations. We believe that including mixed correlators will be rewarding in many other bootstrap studies that are currently ongoing.

Specifically, for $O(N)$ symmetric CFTs, we found that the scaling dimensions of the lowest $O(N)$ vector scalar $\phi$ and $O(N)$ singlet scalar $s$ are constrained to lie in a closed region in the $(\De_\phi,\De_s)$ plane. Our assumptions, besides conformal and $O(N)$ symmetry, were crossing symmetry, unitarity, and --- crucially --- the absence of other relevant scalars in the $O(N)$ singlet and vector sectors. This is completely analogous to the $\mathbb{Z}_2$-symmetric case where similar assumptions isolate a small allowed region around the Ising model in the $(\De_\s,\De_\e)$ plane.  Our allowed regions represent rigorous upper and lower bounds on dimensions in the $O(N)$ models. In principle, this approach could be used to compute the scaling dimensions of the $O(N)$ models to a very high precision, assuming that the allowed region will shrink to a point with increased computational power. However, our results suggest that the region either does not shrink to a point, or the convergence is slow in the present setup. Therefore, our uncertainties are currently larger than the error bars obtained using other methods.\footnote{If one is willing to assume that the $O(N)$ models live near kinks in our allowed regions, then more precise determinations are possible.} In particular, we have not yet resolved the discrepancy between Monte Carlo simulations and experiment for the value of $\De_s$ in the $O(2)$ model.

Including more correlators could result in significantly improved bounds on operator dimensions. In the case of $O(N)$ symmetric CFTs, it would be natural to include the lowest dimension $O(N)$ symmetric tensor as an external operator in the crossing equations. In the $O(N)$ models, this operator actually has a lower dimension than $s$. This is an important difference from the Ising model, where $\phi$ and $s$ are the two lowest dimensional scalars in any sector of the theory. Our present bounds on the lowest symmetric tensor treated it as an internal operator in the crossing equations. Including it as an external operator would open access to many other $O(N)$ representations. Perhaps the $O(N)$ models are not uniquely determined by the condition of only one relevant $O(N)$ singlet and vector scalar, and we must also specify something about these other representations. Studying the $O(N)$ models in other dimensions (such as in 5d~\cite{Nakayama:2014yia,Bae:2014hia,Chester:2014gqa,Fei:2014yja,Fei:2014xta}) may also help to shed light on these issues. We plan to further explore these questions in the future.

In addition to scaling dimensions, it is also important to determine OPE coefficients. Here we presented an example in the computation of the current central charge $C_J$. In the case of $O(2)$ symmetry, this yields the current most precise prediction for the high-frequency conductivity in $O(2)$-symmetric systems at criticality. It will be interesting to extend these mixed-correlator computations to other OPE coefficients in the $O(N)$ models such as the stress-tensor central charge $C_T$ and 3-point coefficients appearing in $\<JJs\>$ and $\<JJT\>$ which control frequency-dependent corrections to conductivity. Pursuing the latter will require implementing the bootstrap for current 4-point functions, a technical challenge for which efforts are ongoing in the bootstrap community.

More generally, the results of this work make it seem likely that scaling dimensions in many other strongly-interacting CFTs can be rigorously determined using the multiple correlator bootstrap. It will be interesting to study mixed correlators in 3d CFTs with fermions and gauge fields -- it is plausible that similar islands can be found for the 3d Gross-Neveu models and 3d Chern-Simons and gauge theories coupled to matter. In 4d, we hope that by pursuing the mixed correlator bootstrap we will eventually be able to isolate and rigorously determine the conformal window of QCD. It also be interesting to apply this approach to theories with conformal manifolds to see the emergence of lines and surfaces of allowed dimensions; a concrete application would be to extend the analysis of~\cite{Beem:2013qxa,Alday:2014qfa} to mixed correlators and pursue a rigorous study of the dimension of the Konishi operator in $\cN=4$ supersymmetric Yang-Mills theory at finite $N$. The time is ripe to set sail away from our archipelago and explore the vast ocean of CFTs!

\section*{Acknowledgements}
We thank Chris Beem, Sheer El-Showk, Luca Iliesiu, Emanuel Katz, Igor Klebanov, Daliang Li, Miguel Paulos, Silviu Pufu, Leonardo Rastelli, Slava Rychkov, Subir Sachdev, Erik S{\o}rensen, Andreas Stergiou, Balt van Rees, William Witczak-Krempa, and Ran Yacoby for discussions. We additionally thank the organizers of the PCTS workshop ``Higher Spin Symmetries and Conformal Bootstrap" for facilitating discussions related to this work. The work of DSD is supported by DOE grant number DE-SC0009988 and a William D. Loughlin Membership at the Institute for Advanced Study.  The work of DP and FK is supported by NSF grant 1350180. The computations in this paper were run on the Bulldog computing clusters supported by the facilities and staff of the Yale University Faculty of Arts and Sciences High Performance Computing Center, on the Hyperion computing cluster supported by the School of Natural Sciences Computing Staff at the Institute for Advanced Study and on the CERN cluster.

\appendix

\section{Implementation Details}
\label{sec:appA}

As described in \cite{Kos:2014bka}, the problem of finding $\alpha$ satisfying (\ref{eq:functionalconditions}) can be transformed into a semidefinite program. Firstly, we must approximate derivatives of $\vec V_S, \vec V_T, \vec V_A$, and $\vec V_V$ as positive functions times polynomials in $\Delta$. We do this by computing rational approximations for conformal blocks using the recursion relation described in \cite{Kos:2014bka}.  Keeping only the polynomial numerator in these rational approximations, (\ref{eq:functionalconditions}) becomes a ``polynomial matrix program" (PMP), which can be solved with \texttt{SDPB} \cite{Simmons-Duffin:2015qma}.

Three choices must be made to compute the PMP.  Firstly, $\kappa$ (defined in appendix A of \cite{Simmons-Duffin:2015qma}) determines how many poles to include in the rational approximation for conformal blocks.  Secondly, $\Lambda$ determines which derivatives of conformal blocks to include in the functionals $\alpha$.  Specifically, we take
\be
\alpha_i(F) &=& \sum_{m+n\leq \Lambda} a_{imn}\ptl_z^m \ptl_{\bar z}^n F(z,\bar z)|_{z=\bar z = \frac 1 2}.
\ee
Some of these derivatives vanish by symmetry properties of $F$.  The total number of nonzero components of $\vec\alpha$ is
\be
\dim(\vec \alpha) &=& 2\frac{\lfloor \frac{\Lambda+2}{2}\rfloor(\lfloor \frac{\Lambda+2}{2}\rfloor+1)}{2} + 5\frac{\lfloor \frac{\Lambda+1}{2}\rfloor(\lfloor \frac{\Lambda+1}{2}\rfloor+1)}{2}.
\ee
Finally, we must choose which spins to include in the PMP.  The number of spins depends on $\Lambda$ as follows
\be
\label{eq:spinsets}
S_{\L=19} &=& \{0,\dots,26\} \cup \{49,50\},\nn\\
S_{\L=27} &=& \{0,\dots,26\} \cup \{29,30,33,34,37,38,41,42,45,46,49,50\},\nn\\
S_{\L=35} &=& \{0,\dots,44\} \cup \{47,48,51,52,55,56,59,60,63,64,67,68\},\nn\\
S_{\L=39} &=& \{0,\dots,54\} \cup \{57,58,61,62,65,66,69,70,73,74,77,78\}.
\ee

We use \texttt{Mathematica} to compute and store tables of derivatives of conformal blocks.  Another \texttt{Mathematica} program reads these tables, computes the polynomial matrices corresponding to the $\vec V$'s, and uses the package \texttt{SDPB.m} to write the associated PMP to an \texttt{xml} file.  This \texttt{xml} file is then used as input to \texttt{SDPB}.  Our settings for \texttt{SDPB} are given in table \ref{tab:parameters}.

\begin{table}[!htb]
\centering
\begin{tabular}{|l|c|c|c|c|}
\hline
$\Lambda$ & 19 & 27 & 35 & 39 \\
$\kappa$ & 14 & 20 & 30 & 36 \\
spins & $S_{\L=19}$ & $S_{\L=27}$ & $S_{\L=35}$ & $S_{\L=39}$\\
\texttt{precision} & 448 & 576 & 768 & 896\\
\texttt{findPrimalFeasible} & True & True & True & True\\
\texttt{findDualFeasible} & True & True & True & True\\
\texttt{detectPrimalFeasibleJump} & True & True & True & True\\
\texttt{detectDualFeasibleJump} & True & True & True & True\\
\texttt{dualityGapThreshold} & $10^{-30}$ & $10^{-30}$ & $10^{-30}$ & $10^{-70}$ \\
\texttt{primalErrorThreshold} & $10^{-30}$ & $10^{-30}$ & $10^{-40}$ & $10^{-70}$ \\
\texttt{dualErrorThreshold} & $10^{-30}$ & $10^{-30}$ & $10^{-40}$ & $10^{-70}$ \\
\texttt{initialMatrixScalePrimal} & $10^{40}$ & $10^{50}$ & $10^{50}$ & $10^{60}$\\
\texttt{initialMatrixScaleDual} & $10^{40}$ & $10^{50}$ & $10^{50}$ & $10^{60}$\\
\texttt{feasibleCenteringParameter} & 0.1 & 0.1 & 0.1 & 0.1 \\
\texttt{infeasibleCenteringParameter}  & 0.3 & 0.3 & 0.3 & 0.3\\
\texttt{stepLengthReduction} & 0.7 & 0.7 & 0.7 & 0.7\\
\texttt{choleskyStabilizeThreshold}  & $10^{-40}$ & $10^{-40}$ & $10^{-100}$ & $10^{-120}$ \\
\texttt{maxComplementarity} & $10^{100}$ & $10^{130}$ & $10^{160}$ & $10^{180}$\\
\hline
\end{tabular}
\caption{\texttt{SDPB} parameters for the computations of scaling dimension bounds in this work. For $C_J$ bounds we need to set all of the Boolean parameters in the table to False. In addition to that, we used $\texttt{dualityGapThreshold}=10^{-10}$, while all the rest of the parameters were kept at the same values as for the dimension bounds.}
\label{tab:parameters}
\end{table}

Finally let us conclude with some comments on the precision of the plots presented in the main text.
Conformal blocks of correlation functions involving operators of nonequal dimensions depend nontrivially on the difference of the dimensions. Hence, when computing the boundary of various allowed regions, it is convenient to perform a scan over a lattice of points. The vectors generating the lattice points are shown in table~\ref{tab:plotsprecision}. The smooth regions shown in figs. \ref{fig:Archipelago}, \ref{fig:O2island}, \ref{fig:O3island}, and \ref{fig:O4island} are the results of a least-squares fit, subject to the constraint that allowed lattice points should lie inside the curves while excluded ones lie outside. In table~\ref{tab:plotsprecision} we also show the maximal perpendicular distance of these points to the curves.

The bounds on $C_J$ shown in figures \ref{fig:O2CJ} and \ref{fig:O3CJ} were computed for the lattices of points that were found to be allowed in figures \ref{fig:O2island} and \ref{fig:O3island}. For each point on the lattice, the bound on $C_J$ was determined to a precision of $10^{-10}$. The smooth regions were obtained by interpolation and the maximum distance of the computed points to the boundry of the shaded region is again reported in table~\ref{tab:plotsprecision}.

\begin{table}[!htb]
\centering
\begin{tabular}{|c|c|c|c|c|c|}
\hline
& & allowed & excluded& $v_1$ & $v_2$ \\
\hline
&$\L=19$ & 0.00025 & 0.00060 & $(10^{-4},10^{-4})$&$(0,10^{-3})$\\
$O(2)$&$\L=27$& 0.000084 & 0.00025  & $(10^{-4},10^{-4})$&$(0,4\cdot10^{-4})$\\
&$\L=35$& 0.00021 & 0.00062 & $(5\cdot10^{-5},5\cdot10^{-5})$&$(0,4\cdot10^{-4})$\\
\hline
&$\L=19$  & 0.00043 & 0.0020 & $(10^{-4},10^{-4})$&$(0,2\cdot10^{-3})$\\
$O(3)$&$\L=27$& 0.00044 & 0.0019  & $(10^{-4},10^{-4})$&$(0,2\cdot10^{-3})$\\
&$\L=35$& 0.00041 & 0.0013 & $(10^{-4},10^{-4})$&$(0,10^{-3})$\\
\hline
&$\L=19$ & 0.00040 & 0.00041  & $(10^{-4},10^{-4})$&$(0,2\cdot10^{-3})$\\
$O(4)$&$\L=27$& 0.00048 & 0.00048  & $(10^{-4},10^{-4})$&$(0,2\cdot10^{-3})$\\
&$\L=35$ & 0.00029 & 0.00062  & $(10^{-4},10^{-4})$&$(0,2\cdot10^{-3})$\\
\hline
$O(20)$&$\L=35$ & 0.00014 & 0.00023  & $(10^{-4},10^{-4})$&$(0,2\cdot10^{-3})$\\
\hline
$O(2)$: $C_J$&$\L=27$ & 0.00005 &  -  & $(10^{-4},10^{-4})$&$(0,\cdot10^{-3})$\\
\hline
$O(3)$: $C_J$&$\L=27$ & 0.0001 &  -  & $(10^{-4},10^{-4})$&$(0,2\cdot10^{-3})$\\
\hline
\end{tabular}
\caption{Maximal distance between the computed allowed and excluded points and the curves shown in figs. \ref{fig:Archipelago}, \ref{fig:O2island}, \ref{fig:O3island}, \ref{fig:O4island}, \ref{fig:O2CJ} and \ref{fig:O3CJ}. The vectors $v_1$ and $v_2$ describe the direction and spacing of the computed grids in the $(\Delta_{\phi},\Delta_s)$ plane. For the $C_J$ bounds we use the same lattices in the $(\Delta_{\phi},\Delta_s)$ plane. The reported maximal distance in the table is the vertical distance of the computed points to the regions shown in the right panels of figures \ref{fig:O2CJ} and \ref{fig:O3CJ}.}
\label{tab:plotsprecision}
\end{table}

\newpage
\section{Symmetric Tensor Scan}
\label{sec:appB}
\begin{figure}[htbp]
\begin{center}
\includegraphics[scale=0.45]{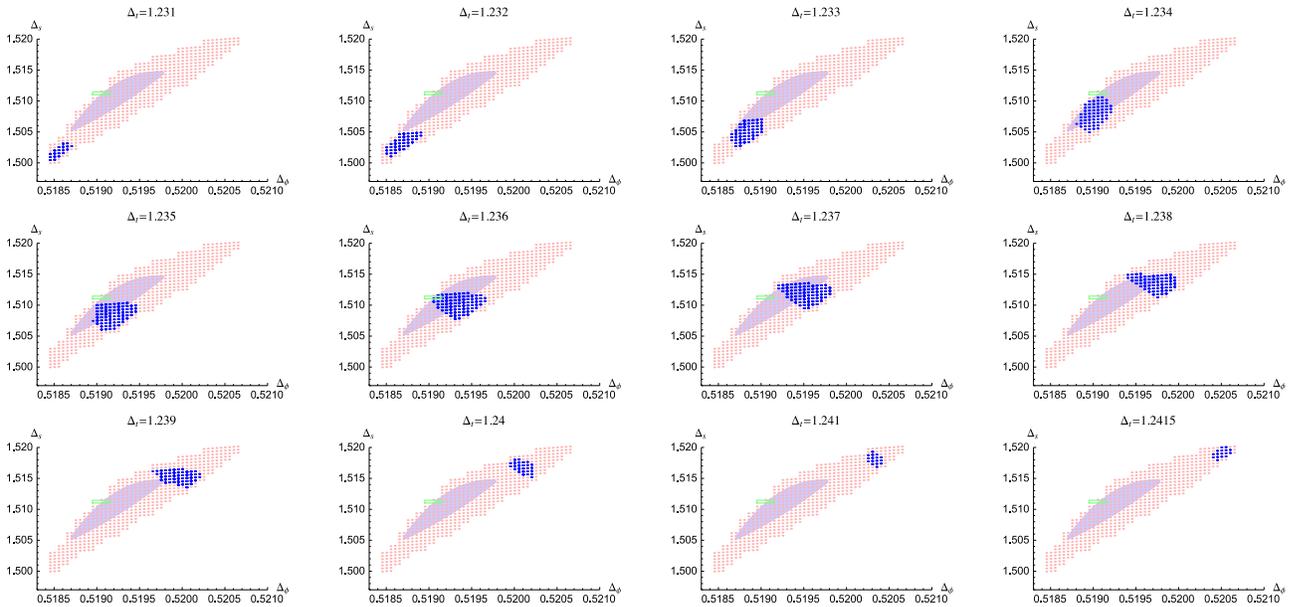}
\caption{Allowed points in the $(\De_\f,\De_s)$ plane for different values of $\De_t$ in $O(2)$ symmetric CFTs at $\Lambda = 19$ (dark blue). The light blue shows the allowed region at $\Lambda=35$ without any assumptions on the symmetric tensor spectrum. The green rectangle is the Monte Carlo estimate~\cite{Campostrini:2006ms}.}
\label{fig:O2scan}
\end{center}
\end{figure}

\begin{figure}[htbp]
\begin{center}
\includegraphics[scale=0.45]{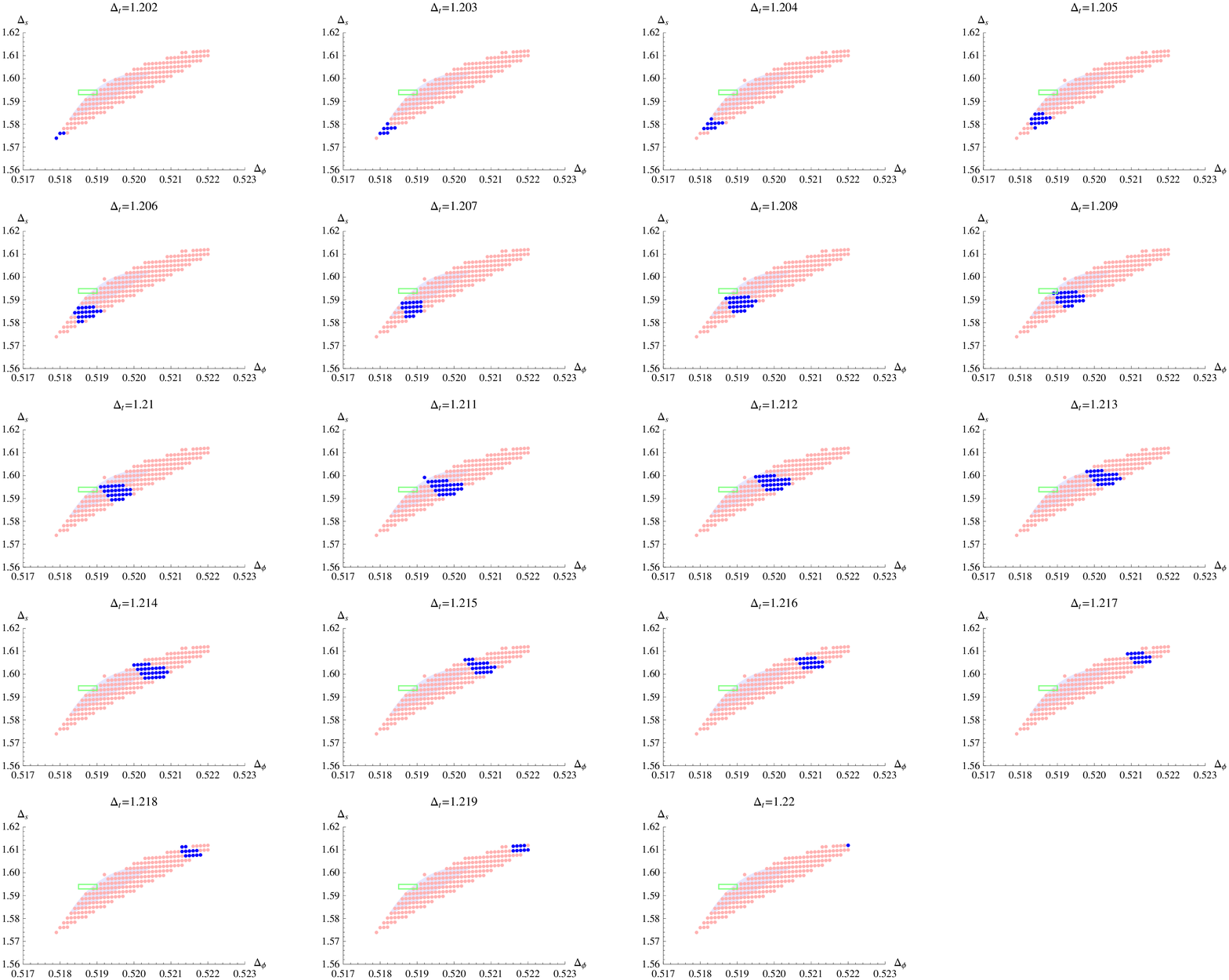}
\caption{Allowed points in the $(\De_\f,\De_s)$ plane for different values of $\De_t$ in $O(3)$ symmetric CFTs at $\Lambda = 19$ (dark blue). The light blue shows the allowed region at $\Lambda=35$ without any assumptions on the symmetric tensor spectrum. The green rectangle is the Monte Carlo estimate~\cite{Campostrini:2002ky}.}
\label{fig:O3scan}
\end{center}
\end{figure}

\begin{figure}[htbp]
\begin{center}
\includegraphics[scale=0.45]{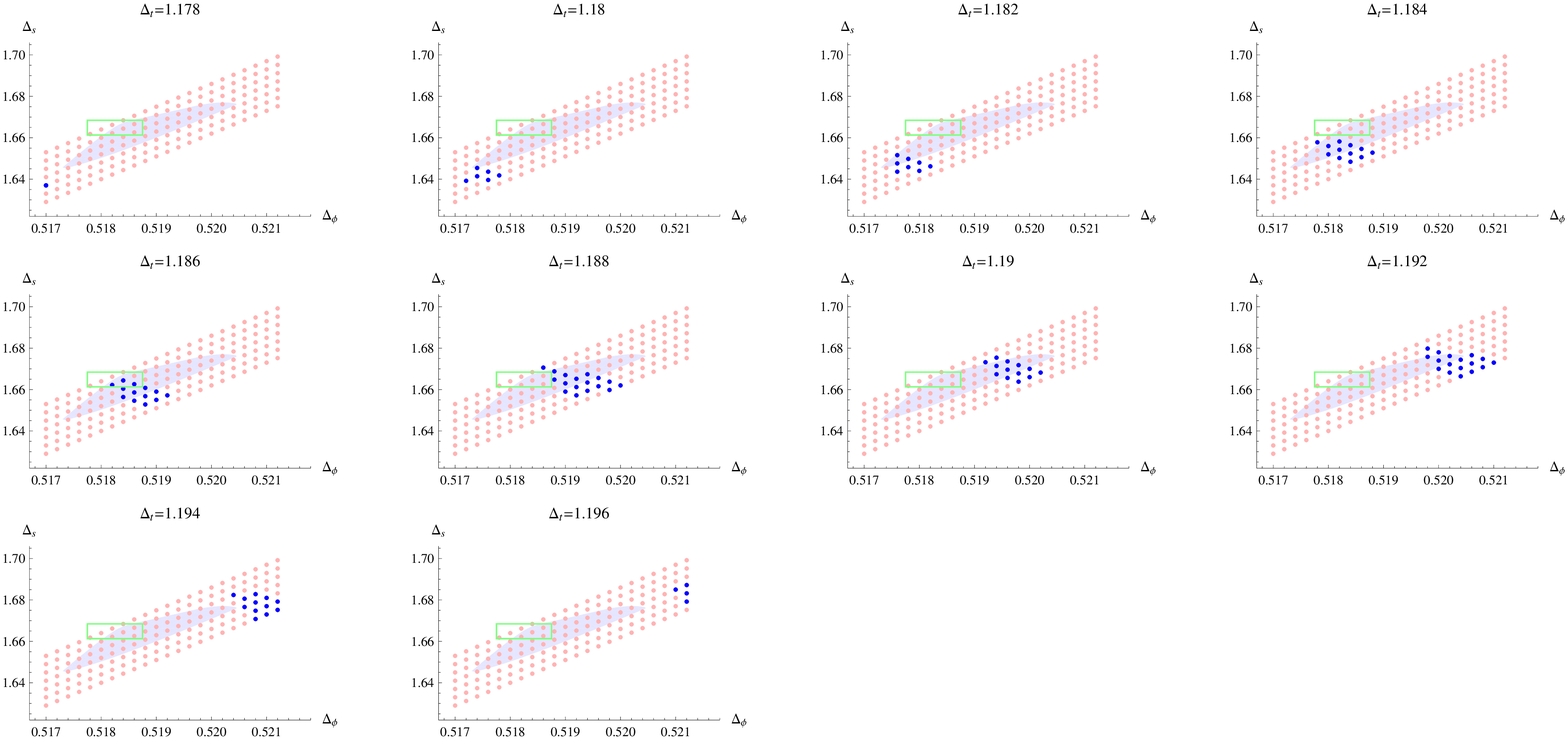}
\caption{Allowed points in the $(\De_\f,\De_s)$ plane for different values of $\De_t$ in $O(4)$ symmetric CFTs at $\Lambda = 19$ (dark blue). The light blue shows the allowed region at $\Lambda=35$ without any assumptions on the symmetric tensor spectrum. The green rectangle is the Monte Carlo estimate~\cite{Hasenbusch:2000ph}.}
\label{fig:O4scan}
\end{center}
\end{figure}

In this appendix we collect some detailed scans of the allowed region of $(\De_\f,\De_s,\De_t)$ space for $O(N)$ models with $N=2,3,4$. The results for the $O(2)$ model are also presented as a 3d plot in figure \ref{fig:O2island3d}. Here we show plots in the $(\De_\f,\De_s)$ plane at fixed values of $\De_t$. The scans for $O(2)$, $O(3)$ and $O(4)$ are shown in figs. \ref{fig:O2scan}, \ref{fig:O3scan}, and \ref{fig:O4scan}, respectively. Blue points represent the allowed region at $\Lambda = 19$. The light blue shaded area is the allowed region at $\Lambda = 35$, but without any assumptions in symmetric tensor sector; those are the same allowed regions shown in figs. \ref{fig:O2island}, \ref{fig:O3island}, and \ref{fig:O4island}. The final allowed regions with the assumptions on $\De_t$ are thus given by the intersections of the dark blue and light blue regions.

Qualitatively the picture is the same for each value of $N$ and we expect that the projections of the 3d plot into the $(\De_\f,\De_s)$ plane will look similar for even higher values of $N$. In particular, the lowest allowed values of $\De_t$ are obtained at the lower left corner of the allowed region in the $(\De_\f,\De_s)$ plane, while the greatest values are obtained at upper right corner of the allowed region. This allows us to find general bounds on $\De_t$ without doing a whole scan over the $(\De_\f,\De_s)$ plane; it is enough to find bounds on $\De_t$ at the corner points.

\clearpage
\bibliography{Biblio}{}
\bibliographystyle{utphys}

\end{document}